\documentclass{ieeeaccess}
    
\usepackage{cite}
\usepackage{booktabs}
\usepackage{xcolor}
\usepackage{multirow}
\usepackage{graphicx,enumitem}
\usepackage[numbers]{natbib}
\usepackage{algorithm}
\usepackage{algorithmic}
\usepackage{amsmath}
\usepackage{amssymb}
\usepackage{dsfont}
\usepackage{array}
\usepackage{booktabs}
\usepackage{threeparttable}
\usepackage{subfigure}
\usepackage{bm}

\newcommand{\revisiondelete}[1]{{}}

\usepackage{enumitem}
\setlist[itemize]{leftmargin=15pt}

\graphicspath{{pic/}}

\begin{document}
\history{Date of publication xxxx 00, 0000, date of current version xxxx 00, 0000.}
\doi{10.1109/ACCESS.2017.DOI}

\title{QA4PRF: A Question Answering based Framework for Pseudo Relevance Feedback}

\author{\uppercase{Handong Ma}\authorrefmark{1*},
\uppercase{Jiawei Hou}\authorrefmark{1*}, \uppercase{Chenxu Zhu}\authorrefmark{1}, \uppercase{Weinan Zhang}\authorrefmark{1}, \uppercase{Ruiming Tang}\authorrefmark{2}, \uppercase{Jincai Lai}\authorrefmark{2}, \uppercase{Jieming Zhu}\authorrefmark{2}, \uppercase{Xiuqiang He}\authorrefmark{2}, \uppercase{and Yong Yu}\authorrefmark{1},
}
\address[1]{Shanghai Jiao Tong University, Shanghai 200240, China}
\address[2]{Huawei Noah's Ark Lab, Shenzhen 518000, China}
\tfootnote{*Handong Ma and Jiawei Hou are co-first authors with equal contributions. This work was supported by NSFC 61772333.}

\corresp{Corresponding author: Yong Yu (e-mail: yyu@apex.sjtu.edu.cn).}

\markboth
{H. Ma \headeretal: QA4PRF: A Question Answering based Framework for Pseudo Relevance Feedback}
{H. Ma \headeretal: QA4PRF: A Question Answering based Framework for Pseudo Relevance Feedback}

\begin{abstract}
 Pseudo relevance feedback (PRF) automatically performs query expansion based on top-retrieved documents to better represent the user's information need so as to improve the search results. Previous PRF methods mainly select expansion terms with high occurrence frequency in top-retrieved documents or with high semantic similarity with the original query. However, existing PRF methods hardly try to understand the content of documents, which is very important in performing effective query expansion to reveal the user's information need. In this paper, we propose a QA-based framework for PRF called QA4PRF to utilize contextual information in documents. In such a framework, we formulate PRF as a QA task, where the query and each top-retrieved document play the roles of question and context in the corresponding QA system, while the objective is to find some proper terms to expand the original query by utilizing contextual information, which are similar answers in QA task. Besides, an attention-based pointer network is built on understanding the content of top-retrieved documents and selecting the terms to represent the original query better. We also show that incorporating the traditional supervised learning methods, such as LambdaRank, to integrate PRF information will further improve the performance of QA4PRF. Extensive experiments on three real-world datasets demonstrate that QA4PRF significantly outperforms the state-of-the-art methods.
\end{abstract}



\maketitle

\section{Introduction}\label{sec:intro}
Query expansion plays a key role in information retrieval as it tries to find proper terms\footnote{In this paper, we use ``words'' and ``terms'' exchangeably when there is no ambiguity.} to revise the original query so as to better represent the user's information need \cite{carpineto2012survey}.
Many methods have been proposed to select expansion terms. Some of them need the relevance scores of documents according to the given query, which are methods of \emph{relevance feedback}, such as Rocchio's algorithm \cite{rocchio1971relevance,salton1990improving}. 
Another branch of methods, known as \emph{pseudo relevance feedback} (PRF) \cite{Xu:1996:QEU:243199.243202}, 
assumes that the top-retrieved documents are \emph{relevant} to the original query, while the others are \emph{irrelevant}. These ``pseudo'' relevant documents are then used to reformulate the original query by expanding new terms. Compared to methods with relevance feedback, PRF methods are more practical in real-world applications, as the ground-truth relevance scores are not always available. 

There are various PRF methods, which can be categorized into relevance-based models~\cite{lavrenko2001relevance, abdul2004umass, Roy2019Discriminative}, divergence-based models~\cite{zhai2001model, lv2014revisiting}, information-based models~\cite{clinchant2010information,montazeralghaem2017term,montazeralghaem2018theoretical}, matrix factorization-based models~\cite{zamani2016pseudo}, supervised learning-based models~\cite{cao2008selecting} and word embedding methods~\cite{roy2016using, kuzi2016query}. Nonetheless, we argue that existing PRF models are insufficient since they only consider the terms of high occurrence frequency in top-retrieved documents or of high semantic similarity to the original query. 
All of them neglect to understand the content of documents in a human-comprehension way, which is indeed very important to perform effective query expansion to reveal the user's underlying information need.
This kind of contextual interaction information should be taken into account to improve accuracy and interpretability when expanding the query. 
For example, a user issues a query ``\textit{How are Oscar winners selected?}''. After the first-round retrieval, the term ``\textit{film}'' appears 53 times in the top 10 retrieved documents, which is much more than other words (except stopwords). As such, most existing PRF models will select ``\textit{film}'' to expand the original query, but the fact is that the term ``\textit{film}'' has nearly no effect of improving the retrieval performance. With analysis on top-retrieved documents, it is easy to find that ``\textit{voter}'' is the best answer as the expansion term for this query, which can increase the mean average precision (MAP) value by about 10$\%$. However, ``\textit{voter}'' only appears 7 times in top 10 retrieved documents. This example shows the importance of understanding the content of top-retrieved documents in the PRF task.

\begin{figure}[t]
  \centering
  \includegraphics[width=0.8\linewidth]{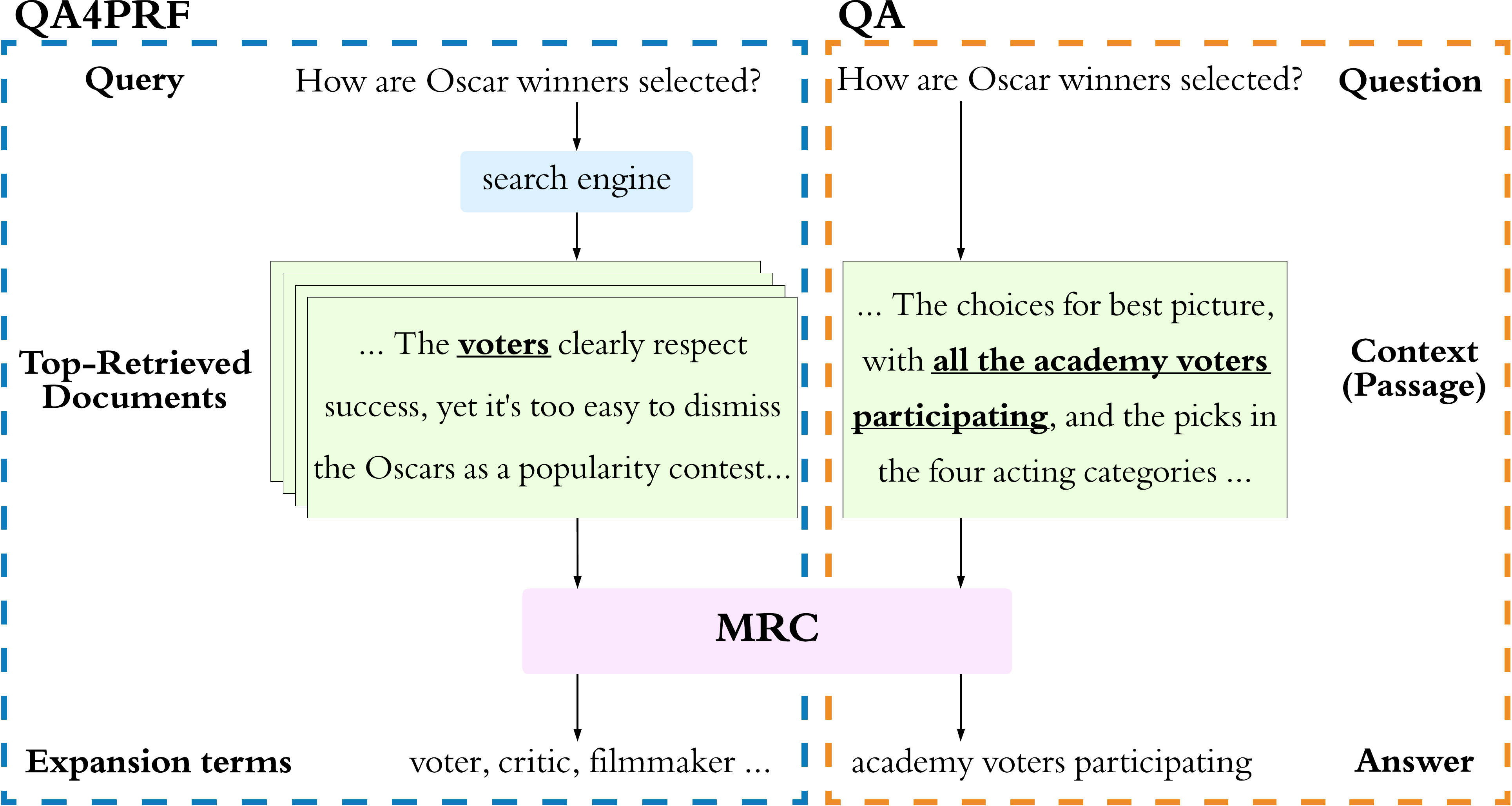}
  \caption{The correspondence between PRF and QA}
  \label{pic:prfqa}
\end{figure}

In the natural language processing field, machine reading comprehension (MRC) \cite{rajpurkar2016squad}, as a framework for question answering (QA) task proposed in 2016, actually provides a high potential method to address this problem. In a QA system, the MRC framework tries to comprehend the question and corresponding passage or contexts and outputs one or several spans of words in the passage as the answer to the question. 
Inspired by MRC, in this paper, we formulate PRF as a QA task: as for PRF, the goal is to find the most effective terms (analogous to the ``answer'' in QA) in each top-retrieved document (analogous to the ``passage'' in QA), for expanding the original query (analogous to the ``question'' in QA). The analogous relationship between PRF and QA is illustrated in Figure~\ref{pic:prfqa}.
With the MRC framework, it is promising to make the PRF model generalize to work on diverse queries and their retrieved documents.

Multi-head attention~\cite{vaswani2017attention} and bi-direction attention~\cite{seo2016bidirectional} are widely used deep learning architectures in QA tasks~\cite{yu2018qanet}, which can capture the global contextual information among long word sequence effectively. As the output of QA is a subset of its input, pointer network~\cite{vinyals2015pointer} and its variants are widely used and shown to be highly effective in MRC frameworks~\cite{qiu2018qa4ie, group2017rnet}. Regarding PRF as a QA task, it is natural to introduce the attention-based pointer network from QA to PRF, aiming to find the most relevant terms from top-retrieved documents for a specific query.

However, applying an attention-based pointer network alone in PRF would neglect some useful terms with high occurrence frequency in the top-retrieved documents. Ignoring such statistical PRF information may lead to query topic drift problems~\cite{raza2018survey}. Therefore, we treat this circumstance, which totally ignores semantic information, as a special case for QA4PRF. To address this issue, we incorporate a supervised learning module\footnote{In our framework, we use LambdaRank~\cite{burges2010ranknet} as the supervised learning method since it is representative and effective learning to rank method with simple implementation. Other supervised learning methods can also be incorporated in our framework without significant modifications.} to estimate the importance of each term from the aspect of statistics, which acts as input features of QA4PRF and yields further improvement of the performance. 

To demonstrate the superiority of our proposed framework, we conduct extensive experiments on three search datasets, where two are public benchmarks, and the other is proprietary. The results show that QA4PRF significantly outperforms the state-of-the-art methods in terms of mean average precision (MAP), normalized discounted cumulative gain (NDCG), and precision at top-retrieved documents. The ablation study further validates the effectiveness of each component of QA4PRF.

To sum up, the main contributions of this work are as follows:
\begin{itemize}
    \item To the best of our knowledge, we are the first to formulate PRF as a QA task and propose a novel QA4PRF framework for query expansion. QA4PRF manages to understand the content of top-retrieved documents to find better expansion terms than existing methods.  
    \item In QA4PRF architecture, an attention-based pointer network is leveraged to learn embedding of each term, considering global contextual information. To further utilize statistical PRF information, we propose to leverage a supervised learning model such as LambdaRank to further enhance the performance of QA4PRF framework.
    \item Extensive experiments on three search datasets demonstrate that QA4PRF achieves significantly better performance than state-of-the-art methods. 
\end{itemize}

The rest of this paper is organized as follows. First, we discuss the related works in Section~\ref{sec:related}. Then, Section~\ref{sec:framework} elaborates the details of the proposed QA4PRF framework. Extensive experiments and results analysis are presented in Section~\ref{sec:exp}. Finally, we conclude this paper in Section~\ref{sec:conclude}.

\section{Related Works}\label{sec:related}

Pseudo relevance feedback (PRF) models are widely used in query expansion and have been shown to be effective~\cite{clinchant2010information, zamani2016pseudo, zhai2001model, lavrenko2001relevance, lv2009comparative, rocchio1971relevance, lucchese2018efficient}. PRF models can be divided into \textit{semantic-based}, \textit{statistics-based} and \textit{hybrid} models according to different sources of input information. In this section, we review these methods separately.

\subsection{Semantic-based PRF models}

Considering the semantic information of query and documents, semantic-based PRF methods adopt word embedding models to generate latent representations of words, and therefore queries and documents. With such latent representation, terms which are most similar to the query are selected for expansion. \citet{roy2016using} proposed to apply kNN based methods to retrieve the most similar terms with respect to a query. \citet{kuzi2016query} utilized the cosine similarity of embeddings to expand the query with terms that are semantically relevant to the query as a whole or to its terms.

Obviously, these embedding methods provide global representations of terms but ignore to comprehend the content of top-retrieved documents, which make it hard for the model to generalize to different queries and top-retrieved documents. In our work, we propose an attention-based pointer network to capture contextual interaction information to address this issue.

\subsection{Statistics-based PRF models}

Statistics-based PRF models assume that the most frequent terms in top-retrieved documents are the best words to expand the query. Such statistical information includes term frequency, inverse document frequency, document length, etc.

\textit{Relevance-based models}~\cite{lavrenko2001relevance, abdul2004umass, Roy2019Discriminative} assume that terms in query are generated by a relevance model $P(w|R)$ (where $R$ denotes the relevance class). RM3~\cite{abdul2004umass} and RM4~\cite{lavrenko2001relevance} provide different approaches to estimate such a relevance model. Based on RM3 model, RM3$^{+}$~\cite{Roy2019Discriminative} takes inverse document frequency of terms into consideration.

\textit{Information-based models}~\cite{clinchant2010information,montazeralghaem2017term,montazeralghaem2018theoretical} select the most informative terms to expand the original query. As in stated \cite{clinchant2010information}, the information of a term in a document can be defined as the statistical difference between situations when the term is in such document and in the whole collection. Based on this, \citet{montazeralghaem2017term} introduced extra term proximity constraints such that a term that appears near a query term, has a higher weight. Recently, \citet{montazeralghaem2018theoretical} raised more interdependence relationships to complete existing constraints.

\emph{Divergence-based models} expand the terms which make the expanded query and relevant documents similar while leading expanded query and the whole collection to be dissimilar. DMM~\cite{zhai2001model} implements this idea through KL-divergence. As a followed up work, MEDMM~\cite{lv2014revisiting} improves DMM by introducing an entropy term as a regularizer, to resolve the skewed feedback issue of DMM.

\emph{Matrix factorization-based methods}~\cite{zamani2016pseudo} treat query expansion as a recommendation problem and establish a document-term weight matrix. Matrix factorization techniques are then used to reformulate the original query by filling the document-term weight matrix.

As can be observed, the aforementioned statistics-based models are all unsupervised learning methods. The work of \citet{cao2008selecting} is the only statistics-based method utilizing \textit{supervised learning} model, which a discriminative model (e.g., support vector machine in this work) is learned to judge whether a term should be chosen for query expansion.

Although these statistical methods can improve the performance of query expansion, all of them totally neglect the contextual interaction information in top-retrieved documents. In our proposed QA4PRF, we formulate PRF as a QA task and apply machine reading comprehension, as a framework for QA tasks, to solve the PRF problem. For the special case which completely ignores any semantic information, we incorporate LambdaRank \cite{burges2010ranknet} to integrate statistical PRF information to improve the performance of our framework.

\subsection{Hybrid PRF models}

Although semantic-based methods can improve retrieval performance after expanding the query, several works~\cite{kuzi2016query, roy2016using} have pointed out that utilizing semantic information alone, such models cannot achieve comparable performance of statistics-based approaches. Due to this observation, a hybrid PRF method is proposed by~\citet{kuzi2016query}, which makes use of both statistical and semantic information. The experiment results show that the semantic information learned by word embedding model improves the performance of RM3~\cite{abdul2004umass} in some cases.

\subsection{Summary} In this paper, we propose a QA based framework for PRF, named as QA4PRF, where PRF is viewed as a QA task. The main differences between our framework and previous works are:
\begin{itemize}
    \item Borrowing idea from QA, an attention-based pointer network is used to learn embedding of each term, capturing contextual interaction information among long word sequence.
    \item To deal with the special circumstance that to utilize statistical PRF information, LambdaRank, a pair-wise learning to rank model with ranked list information, is incorporated to our work reasonably.
\end{itemize}

\section{QA4PRF: Framework and Algorithm}\label{sec:framework}
\subsection{Overview}

Pseudo relevance feedback (PRF) methods are widely adopted in query expansion as it needs no ground-truth relevance scores, which are usually unavailable in industrial IR scenarios. 
In PRF methods, the top-retrieved documents according to a given query are assumed to be to-some-extent relevant. PRF methods select terms from such ``pseudo'' relevant documents, referred as candidate word set, to expand the original query so as to improve the retrieval performance. 
In this work, we recast PRF as a question answering (QA) task to find relevant terms (``answers'' in QA) in each top-retrieved document (``passage'' in QA) for a specific query (``question'' in QA). This section elaborates the details of our proposed QA4PRF framework. The used notations are summarized in Table~\ref{tab:notations} for the ease of presentation.

\begin{table}[h]
	\caption{Notations and descriptions.}
	\label{tab:notations}
	\centering
	\resizebox{0.48\textwidth}{!}{
    	\begin{tabular}{|c|l|}
    		\hline
    		Notation & Description \\
    		\hline
    		$Q$ & The query \\
    		$D_i$ & The $i$-th top-retrieved documents \\
    		$w$ & A term in the candidate word set in general\\
    		$q_i,\ \bm{{q}_i}$ & The $i$-th term and its word embedding in query \\
    		$d_{j,i},\ \bm{{d}_{j,i}}$ & The $i$-th term and its initial embedding in document $D_j$\\
    		$e_i,\ \bm{{e}_i}$ & The $i$-th expansion term and its initial embedding\\
    		$M$ & Number of feedback documents \\
    		$N$ & Number of feedback terms \\
    		$t^Q_w,\ v^Q_w$ & Term frequency and its normalized form of $w$ in query $Q$ \\
    		$t^D_w,\ n^D_w$ & Term frequency and its normalized form of $w$ in document $D$ \\
    		$i_w$ & Inverse document frequency of term $w$\\
    		$C$ & Number of documents in the collection\\
    		$C_w$ & Number of documents contains term $w$\\
    		$\text{avg}_l$ & Average document length \\
    		$|D_i|$ & Length of document $D_i$ \\
    		$\text{FV}(w, Q)$ & Feature vector of term $w$ with respect to query $Q$ \\
    		$W_{\text{QA}}(w)$ & Weight of term $w$ from QA aspect \\
    		$W_{\text{PRF}}(w)$ & Weight of term $w$ leveraging statistical PRF information \\
    		$W(w)$ & Final expansion weight of term $w$ \\
    		$\Theta_z,\ b_z$ & Weight matrices and biases of  hidden layer \\
    		$\beta$ & Feedback coefficient \\
    		$\gamma$ & Trade-off between pointer network (QA aspect) and statistical PRF \\\hline
        \end{tabular}
    }
\end{table}

The pseudo relevance feedback task considered in this paper, is defined as follows. A query consists of $n$ words $Q=\{q_1, q_2\cdots, q_n\}$ and top-M retrieved documents are denoted as $D=\{D_1, D_2,\cdots, D_M\}$, where each document $D_i$ with length $m$ can be represented as $D_i=\{d_{i,1}, d_{i,2},\cdots, d_{i,m}\}$. The output of PRF is a list of terms $E=\{e_1, e_2,\cdots, e_N\}$ from those in the original document set $D$. These N terms are used to expand the original query. In the following, we will use bold letter to denote the \textit{embedding vector} of each term.

In QA4PRF architecture, the attention-based pointer network learns the importance of terms in candidate word set with respect to the query, considering global contextual information. The importance of each term is decided by the semantic relationship between the query and this term, which considers the content of documents at the same time. Then, to utilize statistical PRF information, we construct a feature vector for each word in candidate word set and introduce LambdaRank \cite{burges2010ranknet} as a ranking model to predict importance of each word. Finally, an interpolation method is used, which incorporates the result of LambdaRank to the attention-based pointer network, so as to enhance the performance of QA4PRF. The details of these components are presented in the following subsections.

\subsection{Attention-based Pointer Network}\label{sec:semantic-net}

\begin{figure}[!htb]
    \centering
    \includegraphics[width=0.7\linewidth]{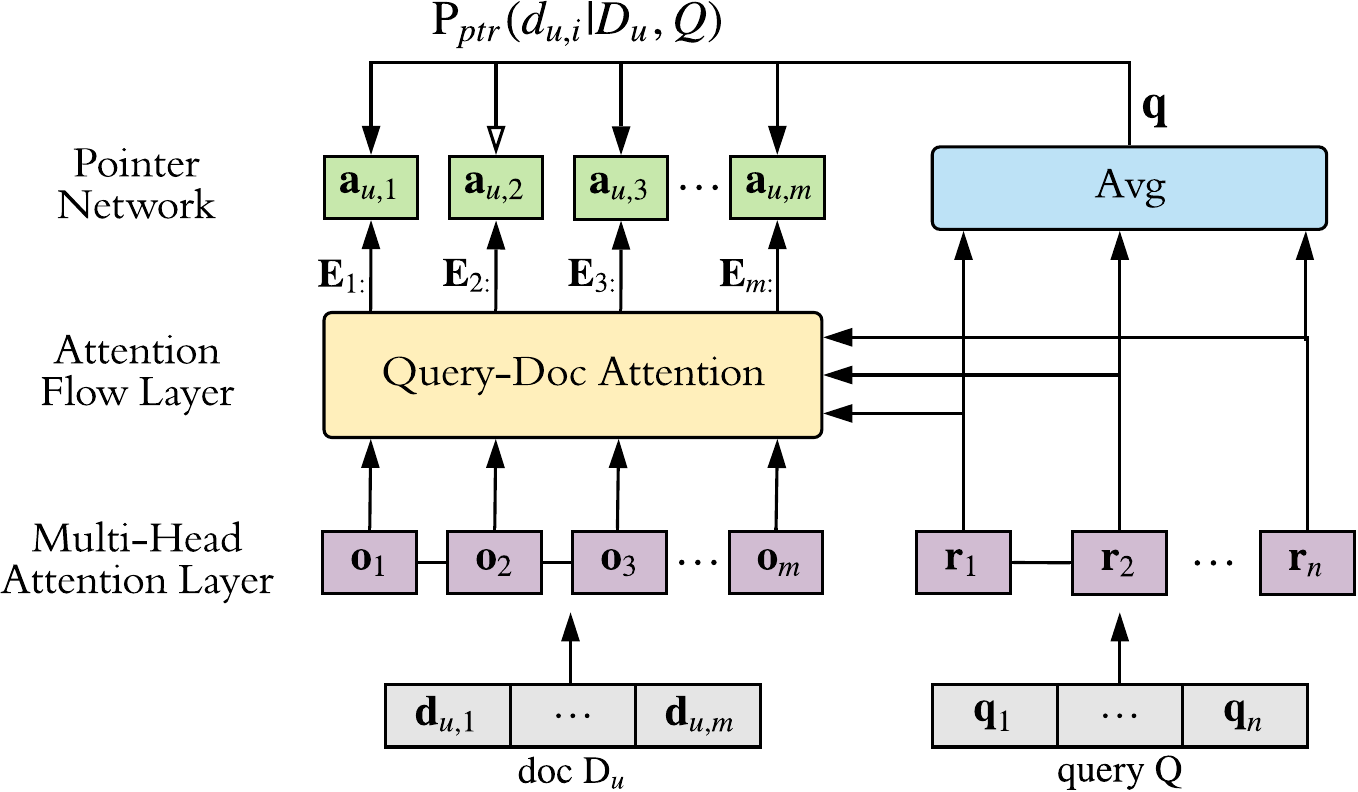}
    \caption{Attention-based pointer network in QA4PRF.}
    \label{pic:specifictrans}
\end{figure}

Instead of generating word embedding with only local co-occurrence relationship in fixed size windows of context in traditional word embedding models, we adapt attention layer to capture the global contextual information among long word sequence efficiently and effectively. Furthermore, a pointer network \cite{vinyals2015pointer} is used to restrict the output of query expansion, which is a set of terms, to be a subset of its input, just like in MRC. The details of attention-based pointer network are described as follows (shown in Figure~\ref{pic:specifictrans}).

\subsubsection{Attention Layer} The attention layer is composed of a multi-head attention layer and an attention flow layer. First, the multi-head attention layer is utilized as the embedding block, which is used by most of the existing MRC models. The inputs of this layer include initial embeddings of terms in a top-retrieved document and the original query. Following~\citet{vaswani2017attention}, the initial embedding of each term is set as the sum of word embedding and positional encoding. The word embedding is initialized from the 300-dimensional FastText vectors \cite{mikolov2018advances}. The positional encoding has the same form with Transformer~\cite{vaswani2017attention}, in order for the model to utilize the order of sequence. Such multi-head attention layer aims to learn embeddings of terms in such a query or document, with the consideration of the relationship between the target word and other words in such a query or document.

The input embeddings can form three matrices $\textbf{Q}$, $\textbf{K}$ (with the number of columns $\text{dim}_K$) and $\textbf{V}$, similar to the Transformer~\cite{vaswani2017attention}. And the output of one attention block can be presented as
\begin{equation}
	\label{equ:attention}
	\text{Attention}(\textbf{Q}, \textbf{K}, \textbf{V})=\text{softmax}\Big(\frac{\textbf{Q}\textbf{K}^T}{\sqrt{\text{dim}_K}}\Big)\textbf{V}.
\end{equation}
The output of multi-head attention is to concatenate the result of each attention block in parallel. To make readers easier to understand, we illustrate our model with a single document $D_u$ and one attention block, where $D_u$ denotes the $u$-th top-retrieved document. In a document attention block, the matrices $\textbf{Q}$, $\textbf{K}$ and $\textbf{V}$ are defined as
\begin{equation}
[\textbf{Q},\textbf{K},\textbf{V}]\ =\ [\textbf{d}_{u,1},\textbf{d}_{u,2},\ldots,\textbf{d}_{u,m}]^T\cdot [\mathbf{W_Q},\mathbf{W_K},\mathbf{W_V}]~,
\end{equation}
where $m$ denotes the length of document $D_u$ for convenience,  $\textbf{d}_{u,1},\textbf{d}_{u,2},\ldots, \textbf{d}_{u,m}$ denote the initial embeddings of terms in document $D_u$, and $\mathbf{W_Q}, \mathbf{W_K}$ and $\mathbf{W_V}$ are the weight matrices. Similar to Equation~\ref{equ:attention}, the embedding block of document $D_u$ is specifically formulated as\footnote{For convenience, we omit weight matrices in the following formulas.}
\begin{equation}
	\label{equ:encoder}
\textbf{t}_{u,i} = \sum\limits_{\textbf{vec}\in\{ \textbf{d}_{u,1},\textbf{d}_{u,2},\ldots,\textbf{d}_{u,m}\}}\text{softmax}\Big(\frac{\textbf{d}_{u, i}^T\cdot \textbf{vec}}{\sqrt{\text{dim}_K}}\Big)\cdot \textbf{vec},
\end{equation}
where $\textbf{t}_{u,i}$ is the embedding of $d_{u,i}$ after the attention block. Then, a two-layer feed-forward network is used, which can be formulated as $\textbf{o}_{u,i}=\text{MLP}(\textbf{t}_{u,i})$. $\text{MLP}$ denotes the feed-forward network and $\textbf{o}_{u,i}$ is the embedding of $d_{u,i}$ after the multi-head attention layer. Here we only show the attention block of document $D_u$ due to space constraints. Such block for query is similar. We represent these embedded vectors of terms $Q=\{q_1, q_2, \ldots, q_n\}$ after the multi-head attention layer as $\{\textbf{r}_1, \textbf{r}_2, \ldots, \textbf{r}_n\}$.

After elaborating the details of multi-head attention, let us present the attention flow layer, which is a Query-Doc attention. In this layer, we compute attention in two directions following~\citet{seo2016bidirectional}. This module is commonly used in many previous machine reading comprehension models such as~\cite{yu2018qanet, chen2017reading}. Such attention block enables each word in query to attend over all words in each top-retrieved document. For convenience, we indicate the input of this block as the document $\textbf{D}_u=[\textbf{o}_1, \textbf{o}_2, \ldots, \textbf{o}_m]$ and the query $\textbf{Q}=[\textbf{r}_1, \textbf{r}_2, \ldots, \textbf{r}_n]$, which are the output of the multi-head attention layer. Firstly, we compute a similarity matrix $\textbf{S}\in\ \mathbb{R}^{m\times n}$ to represent the similarities between each pair of query and document term following~\citet{seo2016bidirectional}. Then we can use matrix $\textbf{S}$ to obtain the attention weights in both directions, namely \textit{Doc-to-Query attention} and \textit{Query-to-Doc attention}.

\textit{Doc-to-Query attention} is utilized to denote which query terms are the most relevant to each word in a top-retrieved document. We can calculate the attention weight by normalizing each row of matrix $\textbf{S}$ by applying the softmax function as $\textbf{a}_{i:}=\text{softmax}(\textbf{S}_{i:})$. The output matrix $\textbf{A}\in \mathbb{R}^{m\times d}$ can be computed as $\textbf{A}_{i:}=\sum_{j}\textbf{a}_{ij}\textbf{r}_j$. Therefore, $\textbf{A}$ contains the attended query vectors for a top-retrieved document.

\textit{Query-to-Doc attention} indicates which term in document is the most relevant to each word in query. So the attention weight can be obtained as $\textbf{b}=\text{softmax}(\text{max}_{\text{column}}(\textbf{S})) \in \mathbb{R}^m$. Then the attended vector matrix of terms in document is $\tilde{b}=\sum_i \textbf{b}_i \textbf{o}_i \in \mathbb{R}^d$. The output maxtrix $\textbf{B}\in \mathbb{R}^{m\times d}$ tiles $\tilde{b}$ by m times.

Finally, the output of Query-Doc attention is computed by applying the average pooling as $\textbf{E} = \text{Avg}(\textbf{A},\textbf{B},\textbf{D}_u^T) \in \mathbb{R}^{m\times d}$.

\subsubsection{Pointer Network} In PRF task, the expansion terms are chosen from the corresponding document, which is the input of PRF model. This is to say, the output of our model is a subset of its input. Due to this reason, we realize a constraint on the output of attention layer with the pointer network, as is often used in QA techniques \cite{wang2017gated}. The probability of expanding each term in the $u$-th top-retrieved document $D_u$ is defined as
\begin{equation}
	\begin{split}
	&a_{u,i} = \Theta_1^T\text{tanh}(\Theta_2\textbf{q} + \Theta_3\textbf{E}_{i:})~,\\
	&P_{\text{pointer}}(d_{u,i}|Q,D_u) = \frac{\text{exp}(a_{u,i})}{\sum_{i'}\text{exp}(a_{u,i'})}~,
	\end{split}
\end{equation}
where $\textbf{q} = \frac{1}{n}\sum_{i=1}^n\textbf{q}_i$ is the embedded vector of the query. Here, $P_{\text{pointer}}(d_{u,i}|Q,D_u)$ means the expansion probability (the output of pointer network) of the $i$-th word in the $u$-th top-retrieved document $D_u$ for a specific query $Q$. 

In the expansion process, the output probability of a candidate term is defined as the summation of the weights of this word in top-$M$ retrieved documents\footnote{Here, $M$ is the number of feedback documents as a hyper-parameter.} by pointer network as
\begin{equation}
	W_{\text{QA}}(w|Q) = \sum\limits^{M}_{u=1}P_{\text{pointer}}(w|Q,D_u)~.
\end{equation}

\subsubsection{Training} To train the attention-based pointer network, the label of expanding word $w$ for query $Q$ needs to be defined. We define the term with the largest $\Delta_{\text{NDCG}}^{Q,w}$ for query $Q$ as ``positive'' word, and others as ``negative'' words, where $\Delta_{\text{NDCG}}^{Q,w}$ represents the NDCG promotion after expanding query $Q$ with term $w$. The network is trained by the cross entropy loss as
\begin{equation}
	\begin{split}
		\mathcal{L} = &-\sum_{i}y(d_{u,i})P_{\text{pointer}}(d_{u,i}|Q,D_u)\\
		&-\sum_i\Big(1-y(d_{u,i})\Big)\Big(1-P_{\text{pointer}}(d_{u,i}|Q,D_u)\Big)~,
	\end{split}
\end{equation}
where $\mathcal{L}$ is the loss of attention-based pointer network and $y(d_{u,i})\in \{0, 1\}$ denotes the label of term $d_{u,i}$. To overcome the difficulty of deep network training, we employ a residual connection and layer normalization in the end of each multi-head attention block following~\citet{vaswani2017attention}.

As mentioned in Section~\ref{sec:intro}, applying MRC framework alone in PRF may neglect statistical PRF information, thereby leading to several problems, such as query topic drift~\cite{raza2018survey}. To handle this issue, we integrate statistical PRF information in QA4PRF to enhance the performance of the framework.

\subsection{Leverage Statistical PRF Information}\label{sec:statistical-ranking}

\subsubsection{Statistics Feature Vector}
In order to leverage statistical PRF information, we construct feature vectors for terms from the aspect of statistics. In PRF, the top-retrieved documents are assumed to be relevant to the original query. The candidate word set includes terms (except stopwords) from top-$M$ retrieved documents of the query. For each term $w$ in the candidate word set, a feature vector of $w$ with respect to query $Q$, i.e., $\text{FV}(w,\ Q)$, is constructed as 
\begin{equation}\label{eq:fv}
	\text{FV}(w,\ Q)\ =\ [v_w^Q,\ i_w,\ n_w^{D_1},\ n_w^{D_2},\ldots,\ n_w^{D_{M}}]^T\ \in\ \mathbb{R}^{M+2}~,
\end{equation}
As presented in Equation~\ref{eq:fv}, the feature vector consists of three factors:
\begin{itemize}
    \item $v_w^Q$ is the normalized term frequency of $w$ in query $Q$, normalized by term frequency summation over all the terms in $Q$ as
    \begin{equation}\label{eq:tf-q}
    v_w^Q = \frac{t^Q_w}{\sum_{q_i\in Q}t^Q_{q_i}}~,
    \end{equation}
    where $t_w^Q$ is the term frequency of $w$ in $Q$.
    \item $i_w$ is the inverse document frequency of $w$ in the whole documents collection:
    \begin{equation}\label{eq:idf-d}
    i_w = \text{log} \Big(\frac{C}{C_w + 1}\Big)~,
    \end{equation}
    where $C$ is the number of documents in the whole collection and $C_w$ is the number of documents containing term $w$ in the whole collection.
    \item $n_w^{D_u}$ is the normalized term frequency of $w$ is a top-retrieved document $D_u$ \cite{clinchant2010information}:
    \begin{equation}\label{eq:tf-idf-d}
    n_w^{D_u} = t^{D_u}_w \text{log}\Big(1 + \alpha\frac{\text{avg}_l}{|D_u|}\Big)~,
    \end{equation}
    where $\text{avg}_l$ is the average document length in the collection and $\alpha$ is a hyper-parameter.
\end{itemize}

Factor $v^Q_w$ and $n^{D_u}_w$ reflect the statistical information of term $w$ in query $Q$ and document $D_u$, while factor $i_w$ reports statistical information of $w$ in the whole collection. Equation~\ref{eq:tf-q} and Equation~\ref{eq:tf-idf-d} consider the length of query and document in different ways. The reason is that queries have almost the same length which is much smaller than documents. This means a small change in length affects the terms in document much less than in query. Therefore, we use average documents length and log function in Equation~\ref{eq:tf-idf-d} to estimate the normalized form.

After generating feature vectors from aforementioned statistical PRF information for each term, we perform supervised learning methods to learn and predict the importance of each word, with respect to the query. In our framework, we apply LambdaRank~\cite{burges2010ranknet} because it is an effective learning to rank method with easy implementation. As stated earlier, other supervised learning methods can also be adopted in our framework without significant modifications.

\subsubsection{LambdaRank} In PRF task, the objective is to improve the retrieval performance after expanding terms. For a specific query $Q$, we aim to improve NDCG by expanding word $w$ with the help of LambdaRank. Such the lift of NDCG after expanding term $w$ is denoted as $\Delta_{\text{NDCG}}^{Q,w}$.

We apply a two-layer neural network to predict the probability of each term in candidate word set to expand as
\begin{equation}
	\text{P}_{\text{lamda}}(w|Q) = \text{sigmoid}\big(\Theta_2\cdot \text{relu}(\Theta_1\cdot \text{FV}(w, Q) + b_1)+ b_2\big)~.
\end{equation}

The training process of LambdaRank in PRF is presented as follows. First, the candidate word set are categorized to a \emph{relevant} word set and an \emph{irrelevant} word set. Similar to the intuition of PRF, relevant word set consists of $N$ words that bring the largest NDCG\footnote{$N$ is the number of query expansion terms, a hyper-parameter which will be discussed in Section~\ref{sec:exp}. Other evaluation metrics such as MAP, ERR are also feasible.} promotion after expansion and irrelevant word set includes the rest. Then, a pair of words $\langle w_i, w_j\rangle$ is selected, such that $w_i$ is selected from relevant word set randomly and $w_j$ is chosen from irrelevant word set randomly. 
That is to say, $\Delta_{\text{NDCG}}^{Q,w_i} > \Delta_{\text{NDCG}}^{Q,w_j}$.
As discussed above, the main objective of PRF is to improve ranking performance of top-retrieved documents. Hence, we use $|\Delta_{\text{NDCG}}^{Q,w_i} - \Delta_{\text{NDCG}}^{Q,w_j}|$ to denote the difference of NDCG promotion when making different choices between $w_i$ and $w_j$. Similar to LambdaRank, we take such difference into consideration in the loss function as
\begin{equation}
	\mathcal{L}_{ij} = |\Delta_{\text{NDCG}}^{Q,w_i} - \Delta_{\text{NDCG}}^{Q,w_j}|\cdot \text{log}\left(1+e^{-\sigma(\text{P}_{\text{lamda}}(w_i|Q)-\text{P}_{\text{lamda}}(w_j|Q))}\right).
\end{equation}

The importance of each term in the candidate word set from the aspect of statistics is defined as ranking score of this term by LambdaRank as
\begin{equation}
	W_{\text{PRF}}(w|Q) = P_{\text{lamda}}(w|Q)~.
\end{equation}

\begin{figure}[t]
	\centering
	\includegraphics[width=0.8\linewidth]{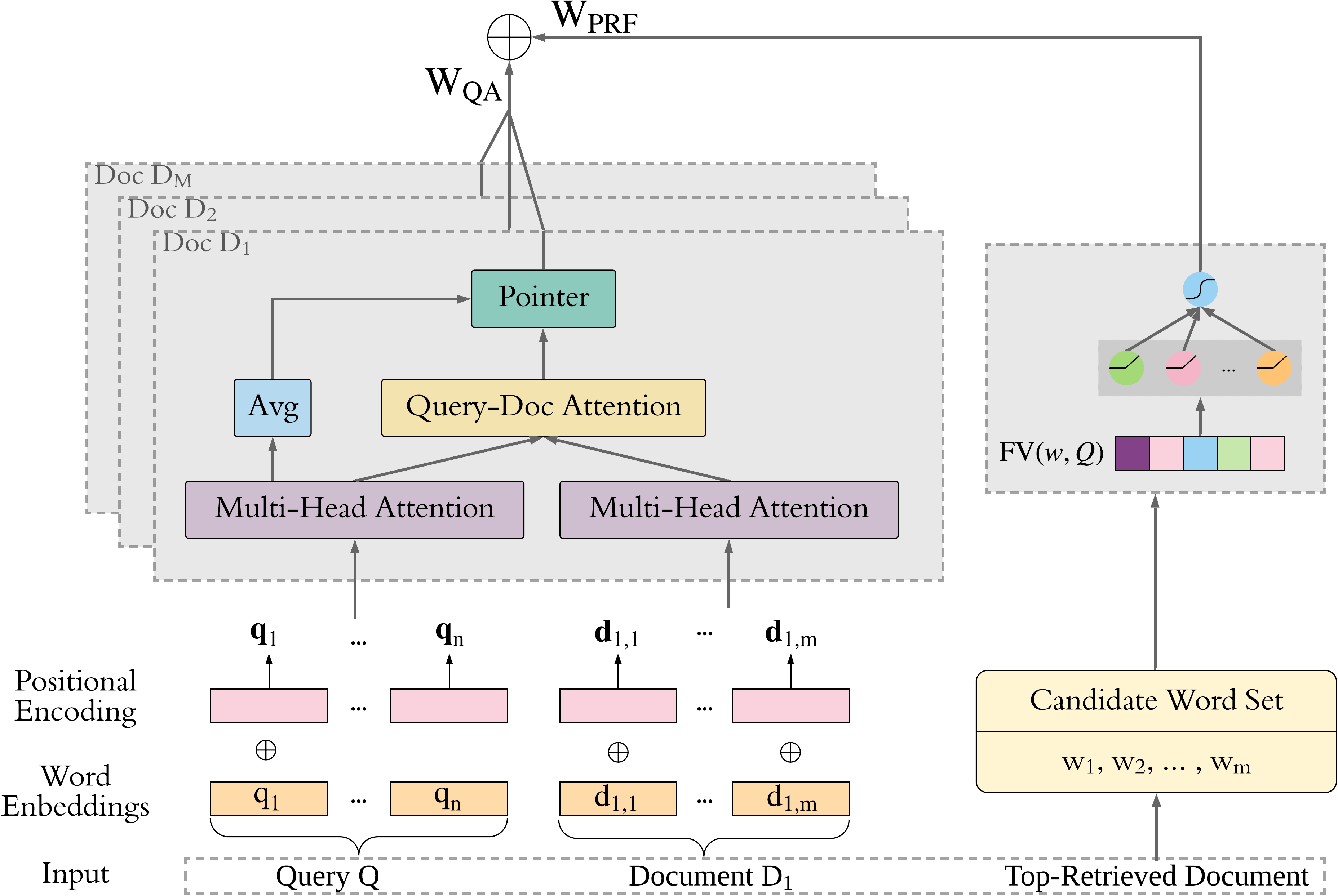}
	\caption{Incorporate LambdaRank to pointer network}
	\label{pic:hybrid}
\end{figure}

\subsection{Final Words Selection}\label{sec:hybrid}

We incorporate the result of LambdaRank to the attention-based pointer network with linear interpolation as shown in Figure~\ref{pic:hybrid}. The weight of term $w_i$ as for query $Q$ can be estimated by QA4PRF framework as
\begin{equation}
	\label{equ:tradeoff}
	W(w_i|Q) = \gamma W_{\text{QA}}(w_i|Q) + (1 - \gamma) W_{\text{PRF}}(w_i|Q)~,
\end{equation}
where $\gamma\in[0,1]$ is a hyper-parameter to trade-off the additional part and the; attention-based pointer network.
According to the weight $W(w_i|Q)$ of each term $w_i$, we sort terms in the candidate word set in descending order and select the top $N$ terms to expand. 

As above mentioned, the query will be expanded by such $N$ terms. To achieve this goal, we define $P(w|Q)$ as the maximum likelihood estimate (MLE) of term $w$ with respect to query $Q$, such that $P(w|Q) = \frac{\text{term frequency of } w}{|Q|}$. The query $Q$ is updated to $Q'$ by expanding term $w$ which selected by weight $W(w|Q)$ as
\begin{equation}
    \label{equ:expansion}
    P(w|Q') = (1 - \beta) P(w|Q) + \beta P(w|E, Q)~,
\end{equation}
where $\beta\in[0,1]$ is feedback coefficient, a hyper-parameter\footnote{Our work focuses on studying how to apply MRC framework to PRF task. We leave some more advanced issues, such as learning $\beta$ for individual terms, in future work.} to make a trade-off between the original query and the expansion terms. $P(w|E, Q)$ is the expansion score of term w for a specific query $Q$. For convenience, we set $P(w|E, Q) = 1$ for $N$ expansion terms and $P(w|E, Q) = 0$ for other words.

\section{Experiments}\label{sec:exp}

In this section, we perform extensive experiments\footnote{Code for our experiments is available at https://bit.ly/2yRvaFr} on three real-world datasets to evaluate our proposed framework. We aim to answer the following research questions (RQs):
\begin{itemize}
	\item \textbf{RQ1:} How does QA4PRF perform as compared with the state-of-the-art PRF models?
	\item \textbf{RQ2:} How do several hyper-parameters (i.e., number of feedback documents, number of feedback terms, feedback coefficient and trade-off of QA) affect the performance of QA4PRF?
	\item \textbf{RQ3:} How do different components of QA4PRF (i.e., attention-based pointer network and LambdaRank) affect its performance?
	\item \textbf{RQ4:} Can our QA4PRF provide expansion results that are easy to interpret?
\end{itemize}

\subsection{Experiment Setup}\label{sec:exp-setup}

\subsubsection{Datasets} 

We conduct experiments on three real-world datasets, where two are public benchmarks and the other is private. \textbf{TREC}\footnote{https://trec.nist.gov/data.html} is an English benchmark dataset. We use the data from TREC robust track 2004 collection. The document collection is from TREC Disks 4 and 5. \textbf{OGeek}\footnote{https://tianchi.aliyun.com/competition/entrance/231688/introduction} data comes from a sub-scenario of OPPO mobile search ranking optimization. This is a Chinese dataset with shorter document length (compared to \textbf{TREC}). Queries are entered by users when searching on mobiles. Documents only contain the title of each page. \textbf{Private} is collected from the user search logs in a mainstream App Store. Queries are entered by users when searching apps on mobiles. Each document is an app in the App Store.

\begin{table}
	\caption{Datasets statistics}
	\label{tab:data-statistics}
	\begin{center}
	\begin{tabular}{|c|c|c|c|c|}
	    \hline
		\textbf{Dataset} & $\#$\textbf{queries} & $\#$\textbf{docs} & \textbf{avg doc length} & $\#$\textbf{labels} \\
		\hline
		TREC & 250 & 174k & 284 & 311k\\
		\hline
		OGeek & 68253 & 662k & 3.18 & 750k\\
		\hline
		Private & 19318 & 79k & 2.82 & 417k \\
		\hline
	\end{tabular}
	\end{center}
\end{table}

To summarize, \textbf{TREC} is a dataset with full documents. The other two are collections with shorter document length. Detailed statistical information of these datasets is shown in Table~\ref{tab:data-statistics}. Following the previous works~\cite{zamani2016pseudo, montazeralghaem2017term}, we only use the title field in \textbf{TREC} to represent each query. All documents are tokenized and stemmed using stemmer with NLTK toolkit~\cite{loper2002nltk} for \textbf{TREC} or Jieba\footnote{https://github.com/fxsjy/jieba} for \textbf{OGeek} and \textbf{Private}. After that, punctuation and stopwords are removed in each document. In all experiments, FastText~\cite{mikolov2018advances} is used to generate initial word embeddings. The pre-trained embedding models can be downloaded from web\footnote{https://fasttext.cc/docs/en/crawl-vectors.html}.
As for retrieval function, we utilize BM25\footnote{The reason why we utilize BM25 as the retrieval function is because our paper focuses on the query expansion model in PRF tasks instead of the retrieval model. BM25 is a simple yet effective method commonly used in previous PRF works~\cite{Roy2019Discriminative, zamani2016pseudo}.}~\cite{robertson1995okapi, robertson1994some}, which is a simple yet effective ranking method in information retrieval. The hyper-parameters of BM25 are decided by cross validation on \textbf{TREC}. As mentioned above, documents in \textbf{OGeek} and \textbf{Private} are much shorter, so we splice top-$M$ documents for more accurate results.

\subsubsection{Baselines} 

As stated in~\cite{diaz2016query, roy2016using}, the performance of word embedding (semantic-based) methods is not effective compared to statistical methods for pseudo relevance feedback. Therefore, we omit semantic-based methods in overall performance comparison, but include them in ablation study to compare their performance with attention-based pointer network of QA4PRF. To compare overall performance, we mainly include statistical and hybrid approaches, totally 11 baselines. We categorize such baselines into different classes, without going into details of how they work (the detailed discussion is presented in Section~\ref{sec:related}).

\begin{itemize}
\item \textbf{NoPRF:} Retrieval by the original query without expansion.
\item \textit{Relevance-based models}: \textbf{RM3}~\cite{abdul2004umass}, \textbf{RM4}~\cite{lavrenko2001relevance} and \textbf{RM3$^+$}~\cite{Roy2019Discriminative} (we choose the best model, which is represented as $\text{RM3}^+_3$ in~\cite{Roy2019Discriminative}).
\item \textit{Divergence-based models}: \textbf{DMM}~\cite{zhai2001model} and \textbf{MEDMM}~\cite{lv2014revisiting}.
\item \textit{Information-based models}: \textbf{LL}~\cite{clinchant2010information}, \textbf{LL(pro)}~\cite{montazeralghaem2017term} and \textbf{LL(ALL)}~\cite{montazeralghaem2018theoretical}.
\item \textit{Matrix factorization-based models}: \textbf{MF}~\cite{zamani2016pseudo}.
\item \textit{Supervised learning-based models}: \textbf{SVM}~\cite{cao2008selecting} (we try to replace SVM by neural networks but it results in degrading the performance. Therefore, we only consider \cite{cao2008selecting} as the baseline in the category).
\item \textit{Hybrid models}: \textbf{RM-Embed}~\cite{kuzi2016query} (we select the best performed model, represented as ``RM-Cent'' in~\cite{kuzi2016query}).
\end{itemize}

\subsubsection{Parameter Setting}
The number of feedback documents $M$, the number of feedback terms $N$, the feedback coefficient $\beta$ and the trade-off of QA $\gamma$ are fine-tuned via 5-fold cross validation. $M$ and $N$ are sweeped in the range $\{$5, 10, 20, ... , 100$\}$, respectively. $\beta$ and $\gamma$ are varied in the scope $\{$0, 0.05, 0.1, ... , 0.95, 1$\}$. For fair comparison, the hyper-parameters of baselines are also determined by cross validation.

\subsubsection{Evaluation Metrics} To evaluate the performance, we leverage mean average precision (MAP), normalized discounted cumulative gain (NDCG) and precision of top-retrieved documents. For \textbf{TREC}, following previous works~\cite{zamani2016pseudo, montazeralghaem2018theoretical}, we select top 1000 documents to evaluate MAP and NDCG, select top 20 documents to evaluate precision. However, for two datasets with shorter documents, \textbf{OGeek} and \textbf{Private}, we report MAP, NDCG and precision of top 5 documents.

To illustrate the robustness of models, we utilize robustness index (RI)~\cite{collins2009reducing} which is defined as $(n_+-n_-)/|Q|$, where $n_+/n_-$ represents the number of queries have better/worse NDCG performance after query expansion and $|Q|$ denotes the total number of test queries. Obviously, a higher RI means more robust.

Furthermore, the Wilcoxon signed-rank test~\cite{wilcoxon1992individual} has been conducted to demonstrate that the differences between our proposed framework and the strongest baselines are significant.

\begin{table*}
	\caption{Overall performance on three datasets}
	\label{tab:overall}
	\resizebox{\textwidth}{!}{
	\begin{threeparttable}
	\begin{tabular}{|c||c|c|c|c||c|c|c|c||c|c|c|c|}
	    \hline
	    \multirow{2}{*}{Method} & \multicolumn{4}{c||}{TREC} & \multicolumn{4}{c||}{OGeek} & \multicolumn{4}{c|}{Private} \\ \cline{2-13}
	    & NDCG & MAP & P@20 & RI & NDCG & MAP & P@5 &RI & NDCG & MAP & P@5 &RI \\
		\hline
		\hline
		No PRF & 0.6084 & 0.2467 & 0.3309 & - & 0.2119 & 0.1678 & 0.0696 & - & 0.5920 & 0.3551 & 0.4320 & - \\
		\hline
		\hline
		RM3 & 0.6348 & 0.2884 & 0.3530 & 0.29 & 0.2255 & 0.1764 & 0.0754 & 0.03 & 0.6052 & 0.3693 & 0.4461 & 0.09 \\
		\hline
		RM4 & 0.6339 & 0.2850 & 0.3490 & 0.27 & 0.2591 & 0.1983 & 0.0893 & \underline{\textbf{0.12}} & 0.6043 & 0.3669 & 0.4436 & 0.05 \\
		\hline
		RM3$^+$ & 0.6379 & 0.2929 & 0.3420 & 0.25 & 0.2450 & 0.1891 & 0.0834 & 0.09 & 0.6050 & 0.3689 & 0.4463 & 0.07 \\
		\hline
		DMM & 0.6236 & 0.2716 & 0.3514 & 0.35 & 0.2619 & 0.2012 & 0.0899 & 0.05 & 0.6053 & 0.3723 & 0.4459 & 0.09 \\
		\hline
		MEDMM & 0.6272 & 0.2747 & 0.3486 & \underline{0.37} & 0.2630 & 0.2020 & \underline{0.0902} & 0.05 & 0.6080 & 0.3732 & 0.4482 & 0.09 \\
		\hline
		LL & 0.6373 & 0.2860 & 0.3445 & 0.20 & 0.2651 & 0.2053 & 0.0899 & 0.10 & \underline{0.6154} & 0.3788 & 0.4533 & \underline{0.11} \\
		\hline
		LL(pro) & 0.6379 & 0.2880 & 0.3444 & 0.17 & \underline{0.2661} & \underline{0.2071} & 0.0896 & 0.10 & 0.6149 & \underline{0.3790} & \underline{0.4557} & 0.08 \\
		\hline
		LL(ALL) & \underline{0.6419} & \underline{0.2956} & 0.3536 & 0.20 & 0.2466 & 0.1914 & 0.0834 & \underline{\textbf{0.12}} & 0.6089 & 0.3714 & 0.4460 & 0.09 \\
		\hline
		MF & 0.6203 & 0.2623 & 0.3428 & 0.33 & 0.2238 & 0.1752 & 0.0747 & 0.03 & 0.5930 & 0.3556 & 0.4342 & 0.01 \\
		\hline
		RM-Embed & 0.6380 & 0.2901 & 0.3504 & 0.29 & 0.2421 & 0.1874 & 0.0821 & 0.06 & 0.6131 & 0.3766 & 0.4513 & 0.10 \\
		\hline
		SVM & 0.6296 & 0.2784 & \underline{0.3542} & 0.31 & 0.2433 & 0.1884 & 0.0825 & 0.06 & 0.6039 & 0.3667 & 0.4437 & 0.02 \\
		\hline
		\hline
		QA4PRF & \textbf{0.6446}$^{*}$ & \textbf{0.2990}$^{*}$ & \textbf{0.3688}$^{*}$ & \textbf{0.43} & \textbf{0.2690}$^{*}$ & \textbf{0.2095}$^{*}$ & \textbf{0.0910}$^{*}$ & 0.11 & \textbf{0.6256}$^{*}$ & \textbf{0.3881}$^{*}$ & \textbf{0.4629}$^{*}$ & \textbf{0.13} \\
		\hline
		Rel. Impr. & 0.42$\%$ & 1.15$\%$ & 4.12$\%$ & - & 1.09$\%$ & 1.16$\%$ & 0.89$\%$ & - & 1.66$\%$ & 2.40$\%$ & 1.58$\%$ & - \\ 
		\hline
	\end{tabular}
	\begin{tablenotes}
	    \normalsize
    	\item * indicates statistically significant improvements (measured by Wilcoxon signed-rank test at $p < 0.05$) over all the baselines.
    	\item ``Rel. Impr.'' presents the relative improvement of our proposed method over the best baseline; ``P@$k$'' represents the precision of top $k$ documents.
	\end{tablenotes}
	\end{threeparttable}}
\end{table*}

\subsection{Overall Performance (RQ1)}\label{sec:exp-overall}

In this subsection, we compare the performance of our proposed QA4PRF with 
several PRF baselines. Table~\ref{tab:overall} reports the overall performance of all models on three datasets, where underlined numbers are the best results of baselines and bold numbers indicate the best results of all models. 

From Table~\ref{tab:overall}, we can conclude that our proposed framework achieves the best performance in the three real-world datasets. Specifically, in these three datasets, compared with the best baseline, QA4PRF obtains the promotion with 0.42\%, 1.09\% and 1.66\% in terms of NDCG (1.15\%, 1.16\% and 2.40\% in terms of MAP, 4.12\%, 0.89\% and 1.58\% in terms of precision), respectively. It demonstrates the superiority of our framework over baselines, in both English and Chinese datasets, with various lengths and numbers of documents. The wilcoxon signed-rank test shows that significant improvement over these metrics are achieved by our method. Besides, the results of RI demonstrate that our proposed QA4PRF is more robust than other baselines in most circumstances.
Further, all the PRF models outperform NoPRF in the three datasets according to table~\ref{tab:overall}, which indicates the effectiveness of PRF methods by using the information in ``pseudo'' relevant documents for query expansion. Information-based models (such as LL, LL(pro), LL(ALL)) perform better than other baselines like relevance-based (e.g., RM3, RM4 and RM3$^+$) and divergence-based (e.g., DMM and MEDMM) under most circumstances. Such findings are consistent with the results and claims in the previous studies~\cite{montazeralghaem2017term, montazeralghaem2018theoretical, clinchant2013theoretical}.

\begin{table}[t]
	\caption{MAP value of diverse categories of queries in TREC}
	\label{tab:map_diff}
	\small
	\begin{center}
	\begin{tabular}{|c||c|c|c|c|c|}
	    \hline
	    \multirow{2}{*}{Method} & \multicolumn{5}{c|}{TREC} \\ \cline{2-6}
		& Bio & Leg & New & Int & Sci \\
		\hline
		$\#$\textbf{queries} & 24$\%$ & 16.4$\%$ & 24.8$\%$ & 11.6$\%$ & 23.2$\%$ \\
		\hline
		\hline
		NoPRF & 0.2693 & 0.2540 & 0.2469 & 0.2084 & 0.2345 \\
		\hline
		RM3$^+$ & 0.3021 & 0.2891 & 0.2978 & 0.2792 & 0.2798 \\
		\hline
		MEDMM & 0.2855 & 0.2706 & 0.2754 & 0.2455 & 0.2690 \\
		\hline
		LL(ALL) & 0.2998 & \textbf{0.3072} & 0.3007 & 0.2809 & 0.2821 \\
		\hline
		SVM & 0.2898 & 0.2791 & 0.2803 & 0.2605 & 0.2706 \\
		\hline
		\hline
		QA4PRF & \textbf{0.3093} & 0.3066 & \textbf{0.3028} & \textbf{0.2858} & \textbf{0.2856} \\
		\hline
	\end{tabular}
	\end{center}
\end{table}

\begin{figure}[t]
	\centering
	\subfigure[Attention-based pointer network]{\includegraphics[width=0.47\linewidth]{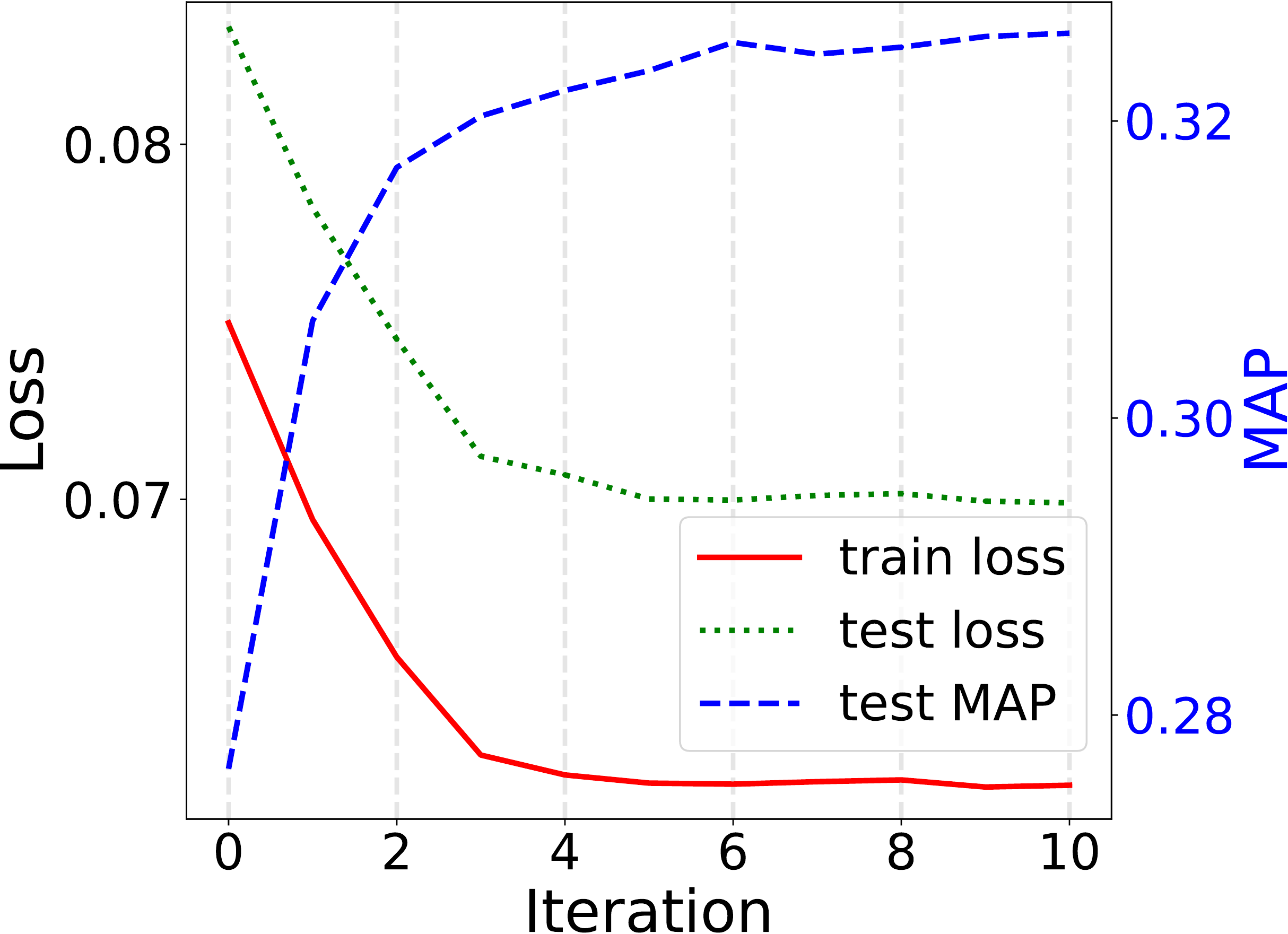}}
	\hspace{.1in}
	\subfigure[LambdaRank]{\includegraphics[width=0.48\linewidth]{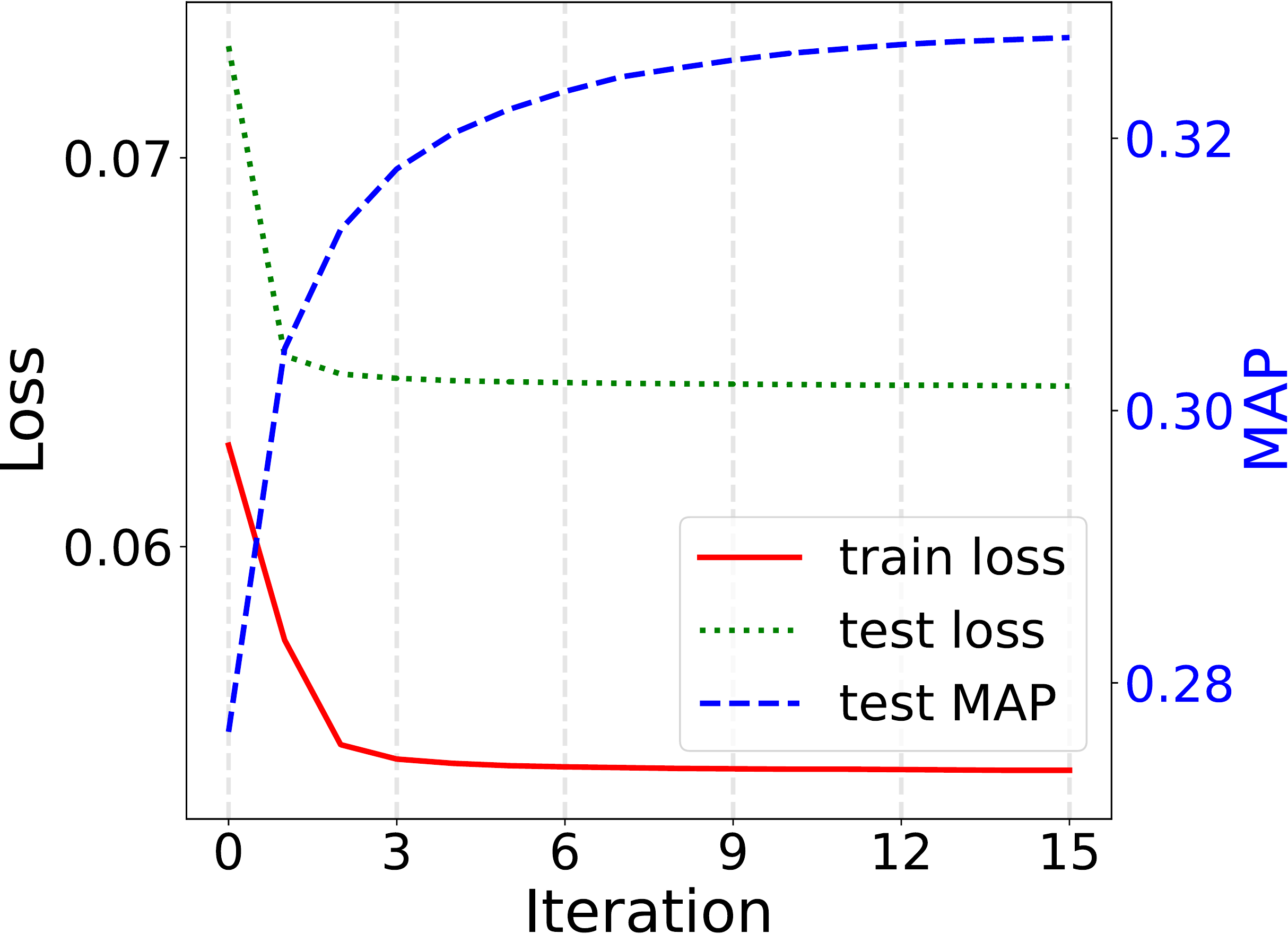}}
	\caption{Training loss, testing loss and testing performance in each iteration on TREC Dataset}
	\label{pic:traincurve}
\end{figure}

Moreover, to further prove the effectiveness of our proposed QA4PRF, we divided queries of TREC into 5 categories (biology and medicine, legal theory, news, international relations, science and technology) according to the user's intent. The specific query classification method is provided in the code. 
Table~\ref{tab:map_diff} shows the result of each category. 
It is obvious that, for most cases, QA4PRF has a 0.7$\%$ to 2.4$\%$ improvement in terms of MAP compared with the best baseline, except for category \textit{legal theory}. Even in the query of \textit{legal theory} category, our proposed model can get almost the same performance as the best baseline. Such results illustrate the superiority of our proposed QA4PRF over baselines for diversified queries.

In addition, to study how QA4PRF performs, we present the training loss, testing loss and testing performance (MAP@1000) in each iteration on TREC in Figure~\ref{pic:traincurve}. Specifically, Figure~\ref{pic:traincurve}(a) shows the training process of attention-based pointer network for the first test set in cross validation, while that of LambdaRank is displayed in Figure~\ref{pic:traincurve}(b). For both methods, we report the best parameter settings. It is obvious that both methods achieve stable performance after about 7 iterations. Extensive studies of these two methods are in Section~\ref{sec:abla}.

\subsection{Hyper-parameter Study (RQ2)}

Our proposed QA4PRF has several key hyper-parameters which may affect the performance of framework, i.e., (i) number of feedback documents $M$, (ii) number of feedback terms $N$, (iii) feedback coefficient $\beta$ and (iv) trade-off $\gamma$ between the attention-based pointer network (QA aspect) and LambdaRank (statistical PRF aspect). In this subsection, to study the impact of these hyper-parameters on our proposed framework, we tune one of them while fixing the others. 

\begin{figure}[t]
	\centering
	\subfigure{
	\begin{minipage}[t]{\linewidth}
		\begin{minipage}[t]{0.45\linewidth}
			\includegraphics[width=\linewidth]{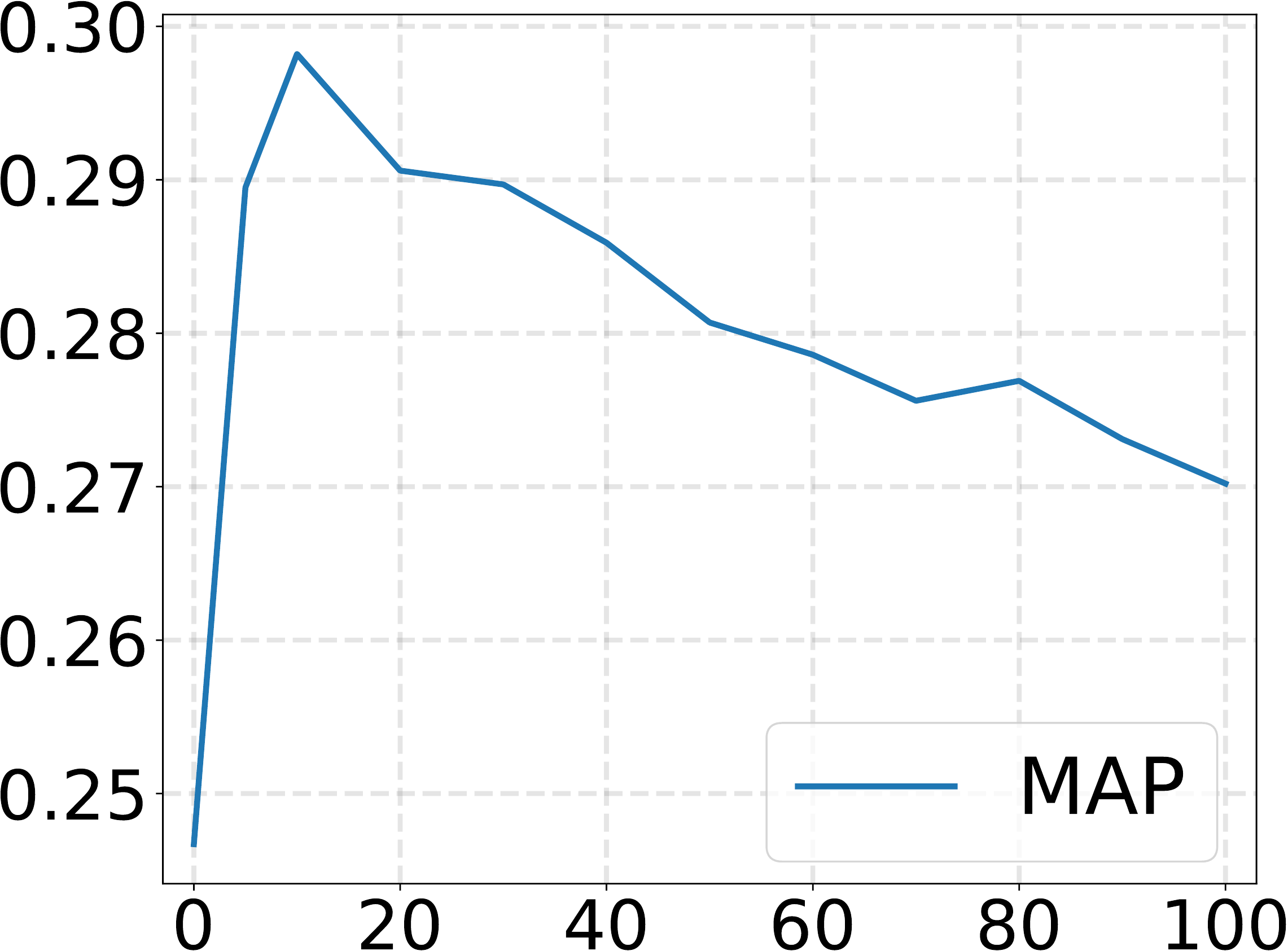}
			\centerline{(a) feedback documents M}
		\end{minipage}%
		\hspace{.2in}
        \begin{minipage}[t]{0.45\linewidth}
			\includegraphics[width=\linewidth]{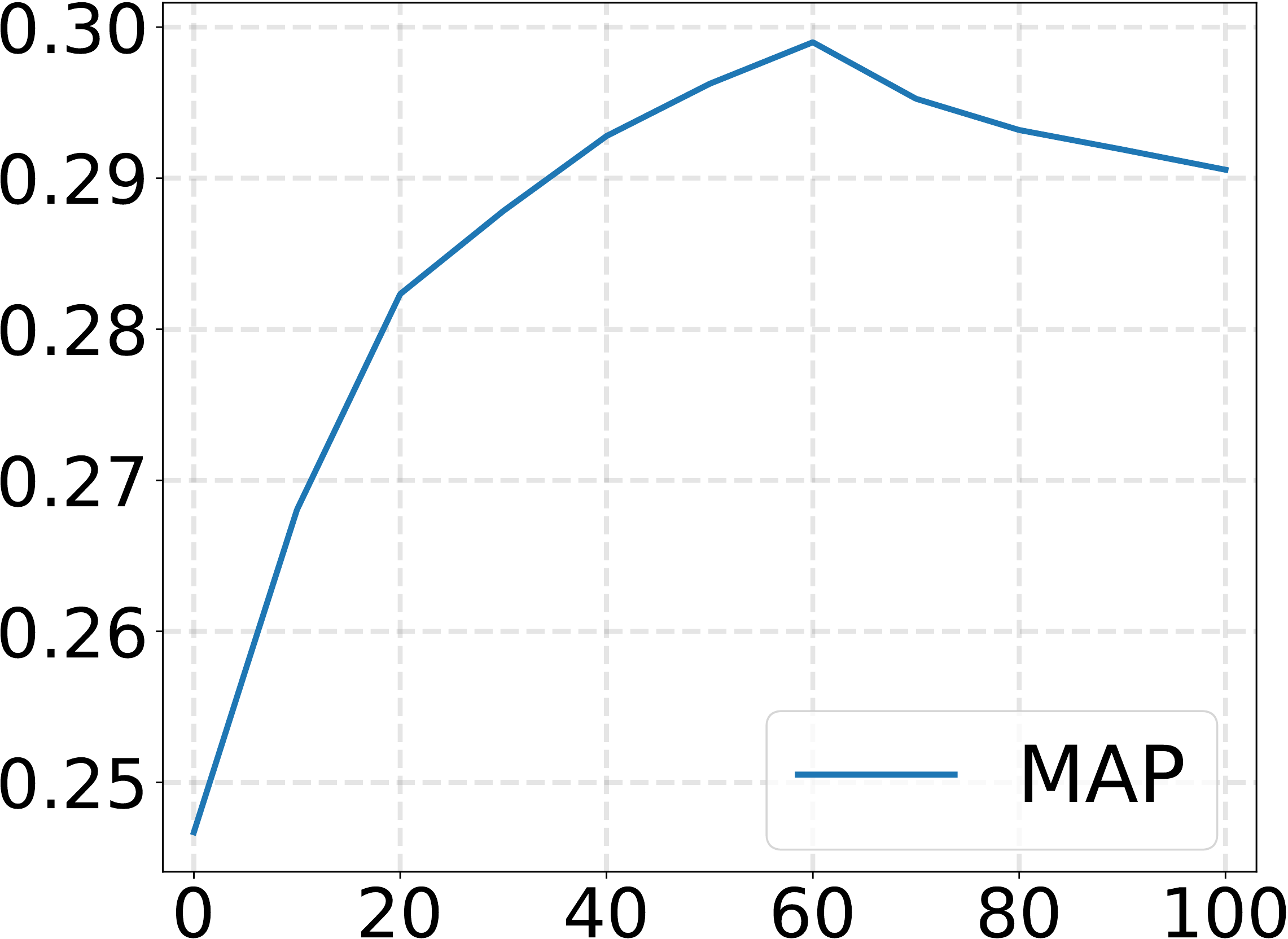}
			\centerline{(b) feedback terms N}
		\end{minipage}%
    \end{minipage}%
	}%

	\subfigure{
	\begin{minipage}[t]{\linewidth}
		\begin{minipage}[t]{0.45\linewidth}
			\includegraphics[width=\linewidth]{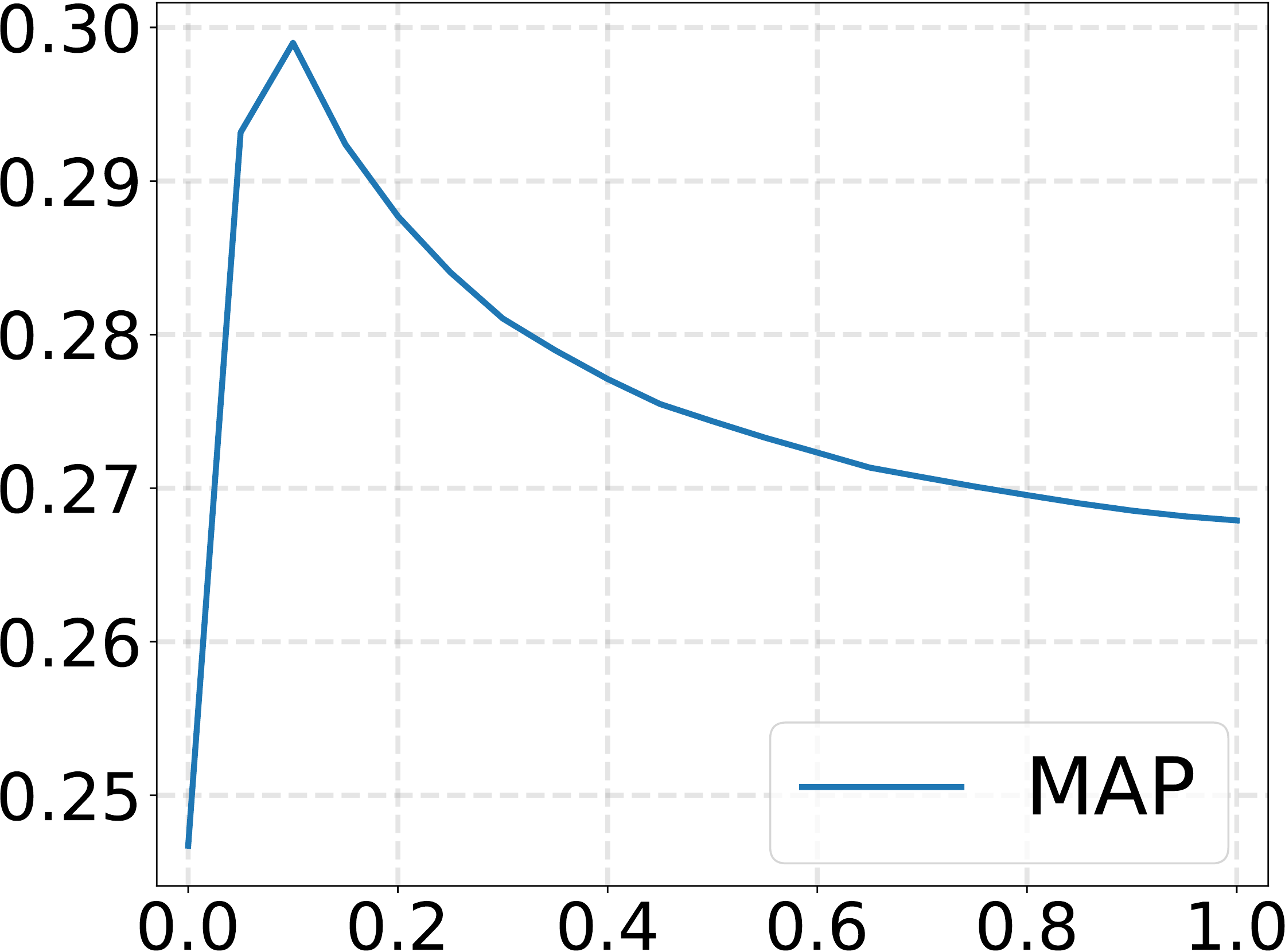}
			\centerline{(c) feedback coefficient $\beta$}
		\end{minipage}%
		\hspace{.2in}
        \begin{minipage}[t]{0.45\linewidth}
			\includegraphics[width=\linewidth]{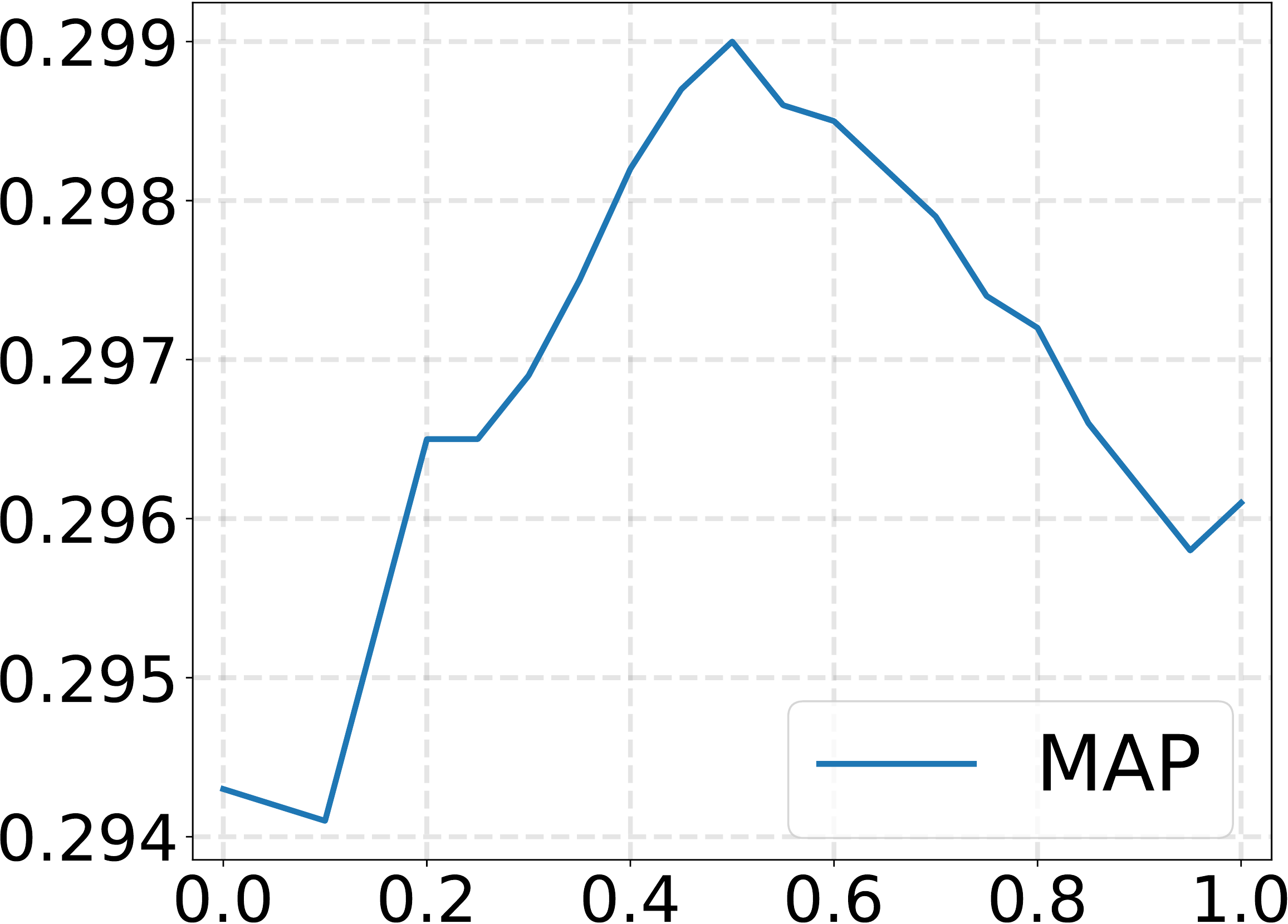}
			\centerline{(d) trade-off of QA $\gamma$}
		\end{minipage}%
    \end{minipage}%
	}%
	\caption{Hyper-parameter study of QA4PRF}
	\label{pic:rq2}
\end{figure}

Specifically, we set the number of feedback documents $M$ (in Section~\ref{sec:statistical-ranking}) as 10 and the number of feedback terms $N$ (in Section~\ref{sec:hybrid}) as 60 which are common settings in the existing PRF methods. For the feedback coefficient $\beta$ (in Equation~\ref{equ:expansion}) and trade-off $\gamma$ (in Equation~\ref{equ:tradeoff}), we fix them as 0.1 and 0.5 by cross validation. Figure~\ref{pic:rq2} presents the experiment results of hyper-parameter study in terms of MAP in TREC dataset. For each hyper-parameter, we have the following observations.

\begin{itemize}
	\item \emph{Number of feedback documents $M$}: The proposed framework performs better when $M$ is enlarged from 5 to 10. The best performance is achieved when $M=10$. When $M$ is larger than 10, the performance of the model keeps dropping as $M$ increases from 10 to 100. This rising-falling phenomenon on the performance is reasonable. When more feedback documents are considered, more terms are included in the candidate word set, so that the chance of expanding query with useful terms is larger, which leads to performance improvement. However, involving too many documents introduces noisy words in the candidate word set, which results in expanding query with useless terms and therefore worse performance.
	\item \emph{Number of feedback terms $N$}: As shown in Figure~\ref{pic:rq2}(b), the MAP trend with the number of feedback terms is rising-falling, which reaches its peak when $N=60$. Expanding more terms increases the chance of formulating useful queries to fit user intent, which helps to boost the performance. However, expanding too many terms results in adding noise to the query and mismatching with user intent, so that the performance is degrading. 
	\item \emph{Feedback coefficient $\beta$}: $\beta$ shows the trade-off between original query and expansion terms. As can be observed in Figure~\ref{pic:rq2}(c), the framework obtains the best performance when $\beta = 0.1$ and the worst performance when $\beta = 0$. Note that $\beta = 0 $ is actually the ranking without query expansion, namely, NoPRF. It validates the effectiveness of query expansion with PRF, which is consistent with an observation in Section~\ref{sec:exp-overall}. When $\beta > 0.1$, MAP value drops slowly, which indicates that the PRF model can help improve retrieval efficiency of the original query, but can not completely replace the original query (i.e., $\beta = 1$).
	\item \emph{Trade-off $\gamma$}: $\gamma$ shows the balance between attention-based pointer network (QA module) and LambdaRank (statistical PRF). It is obvious that the best performance is achieved when $\gamma = 0.5$, which validates that it is effective to considering semantic QA and statistical PRF information simultaneously.
\end{itemize}

\subsection{Ablation Study (RQ3)}\label{sec:abla}
In QA4PRF, there are two components may affect the framework performance: \textit{attention-based pointer network} (in Section~\ref{sec:semantic-net}) which formulates PRF as a QA task and \textit{LambdaRank} (in Section~\ref{sec:statistical-ranking}) which leverages statistical PRF information to enhance the performance. In this subsection, to study the effectiveness of each component, we evaluate the performance of these two components, compared with state-of-the-art baselines over TREC dataset.

To demonstrate the superiority of \textit{attention-based pointer network}, several state-of-the-art semantic-based approaches are chosen as baselines, such as Cent, CombSUM, CombMNZ, CombMAX in~\cite{kuzi2016query} and $k$NN-embed in~\cite{roy2016using}. Besides, we also contain a QA baseline, QANet~\cite{yu2018qanet}, which performs better than other MRC framework. To validate the effectiveness of \textit{LambdaRank}, RM3$^+$, MEDMM, LL(ALL) and SVM are selected as baselines from the ones described in Section~\ref{sec:exp-setup}, as they are the best performed ones. The performance comparison is presented in Table~\ref{tab:abla_sem} and Table~\ref{tab:abla_stat}, respectively. 

\begin{table}[t]
	\caption{Effectiveness of attention-based pointer net}
	\label{tab:abla_sem}
	\small
	\begin{center}
	\begin{tabular}{|c||c|c|c|c|}
	    \hline
	    \multirow{2}{*}{Method} & \multicolumn{4}{c|}{TREC} \\ \cline{2-5}
		& NDCG & MAP & P@20 & RI \\
		\hline
		NoPRF & 0.6084 & 0.2467 & 0.3309 & - \\
		\hline
		\hline
		Cent & 0.6241 & 0.2606 & 0.3355 & 0.19 \\
		\hline
		CombSUM & 0.6251 & 0.2596 & 0.3329 & 0.17 \\
		\hline
		CombMNZ & 0.6243 & 0.2564 & 0.3331 & 0.14 \\
		\hline
		CombMAX & 0.6206 & 0.2554 & 0.3313 & 0.11 \\
		\hline
		kNN-embed & 0.6181 & 0.2557 & 0.3343 & 0.25 \\
		\hline
		QANet & 0.6307 & 0.2800 & 0.3586 & 0.34 \\
		\hline
		\hline
		Atten-pointer & \textbf{0.6418} & \textbf{0.2943} & \textbf{0.3648} & \textbf{0.36} \\
		\hline
	\end{tabular}
	\end{center}
\end{table}

\begin{table}[t]
	\caption{Effectiveness of LambdaRank}
	\label{tab:abla_stat}
	\small
	\begin{center}
	\begin{tabular}{|c||c|c|c|c|}
	    \hline
	    \multirow{2}{*}{Method} & \multicolumn{4}{c|}{TREC} \\ \cline{2-5}
		& NDCG & MAP & P@20 & RI \\
		\hline
		NoPRF & 0.6084 & 0.2467 & 0.3309 & - \\
		\hline
		\hline
		RM3$^+$ & 0.6379 & 0.2929 & 0.3420 & 0.25 \\
		\hline
		MEDMM & 0.6272 & 0.2747 & 0.3486 & 0.37 \\
		\hline
		LL(ALL) & 0.6419 & 0.2956 & 0.3536 & 0.20 \\
		\hline
		SVM & 0.6296 & 0.2784 & 0.3542 & 0.31 \\
		\hline
		\hline
		LambdaRank & \textbf{0.6428} & \textbf{0.2961} & \textbf{0.3587} & \textbf{0.39} \\
		\hline
	\end{tabular}
	\end{center}
\end{table}

One can observe that \textit{Attention-based pointer network} (referred as ``Atten-pointer'' in Table~\ref{tab:abla_sem}) is performed to capture contextual interaction information among long word sequence. The results in Table~\ref{tab:abla_sem} show that such a network achieves much better performance for expanding query than semantic-based baselines. Such results show potential for other QA techniques in PRF task. Compared to QANet, a strong baseline in QA, our attention-based pointer network also gets better performance. This indicates that our model is more suitable for PRF tasks than general QA models.

\textit{LambdaRank} is used to learn the importance of each term based on feature vectors from the aspect of statistics. The results in Table~\ref{tab:abla_stat} show that LambdaRank outperforms all the baselines in terms of NDCG, MAP and precision. Such results demonstrate the effectiveness of our statistical PRF component, compared to the state-of-the-art statistical approaches. Furthermore, as two supervised learning models, the comparison with SVM shows the superiority of applying LambdaRank, which also indicates the potential of trying other learning to rank models in statistical PRF part. Compared with semantic PRF approaches, the effectiveness of statistiscal methods is slightly superior in terms of NDCG and MAP. This observation is consistent with the findings in~\cite{diaz2016query}.

From the above two observations, it can be concluded that solving PRF problem with QA techniques can bring considerable improvement. Incorporating statistical PRF information, at the meantime, to a certain extent can effectively enhance the improvement of our framework.

\begin{table*}[ht]
	\caption{Top 20 expansion terms selscted by LL(ALL) and QA4PRF}
	\label{tab:case}
	\footnotesize
	\begin{center}
	\begin{tabular*}{\textwidth}{ccc}
		\hline
		Original Query & Method & \multicolumn{1}{c}{Expansion Terms} \\
	    \hline
		\multirow{2}{*}{\textbf{Modern Slavery}} & \textbf{LL(ALL)} & \multicolumn{1}{p{0.72\textwidth}}{slaveri, india, gandhi, econom, \textbf{indian}, modern, \textbf{west}, \textbf{life}, societi, tradit, movement, form, time, independ, children, prof, bondag, \textbf{peopl}, \textbf{organ}, nation} \\
		& \textbf{QA4PRF} & \multicolumn{1}{p{0.72\textwidth}}{slaveri, children, econom, india, \textbf{super}, societi, movement, gandhi, \textbf{govern}, prof, \textbf{worker}, \textbf{develop}, nation, independ, bondag, \textbf{labor}, time, modern, tradit, form} \\
		\hline
	    \multirow{2}{*}{\textbf{Diplomatic Expulsion}} & \textbf{LL(ALL)} & \multicolumn{1}{p{0.72\textwidth}}{expuls, diplomat, expel, iranian, soviet, british, iran, embassi, \textbf{foreign}, britain, spi, retali, \textbf{london}, \textbf{titfortat}, \textbf{yesterday}, offici, \textbf{consul}, \textbf{spokesman}, leav, \textbf{countri}} \\
		& \textbf{QA4PRF} & \multicolumn{1}{p{0.72\textwidth}}{iranian, expuls, diplomat, british, retali, \textbf{russian}, expel, spi, offici, \textbf{attach}, \textbf{baghdad}, leav, \textbf{militari}, embassi, soviet, britain, \textbf{union}, iran, \textbf{ministri}, \textbf{iraq}} \\
		\hline
	\end{tabular*}
	\end{center}
\end{table*}


\begin{figure}[ht!]
	\centering
	\subfigure[\normalsize{Top 20 terms assigned by LL(ALL) (the red squares are different expansion terms compared with QA4PRF)}]{
	\begin{minipage}[t]{\linewidth}
		\begin{minipage}[t]{0.49\linewidth}
			\includegraphics[width=\linewidth]{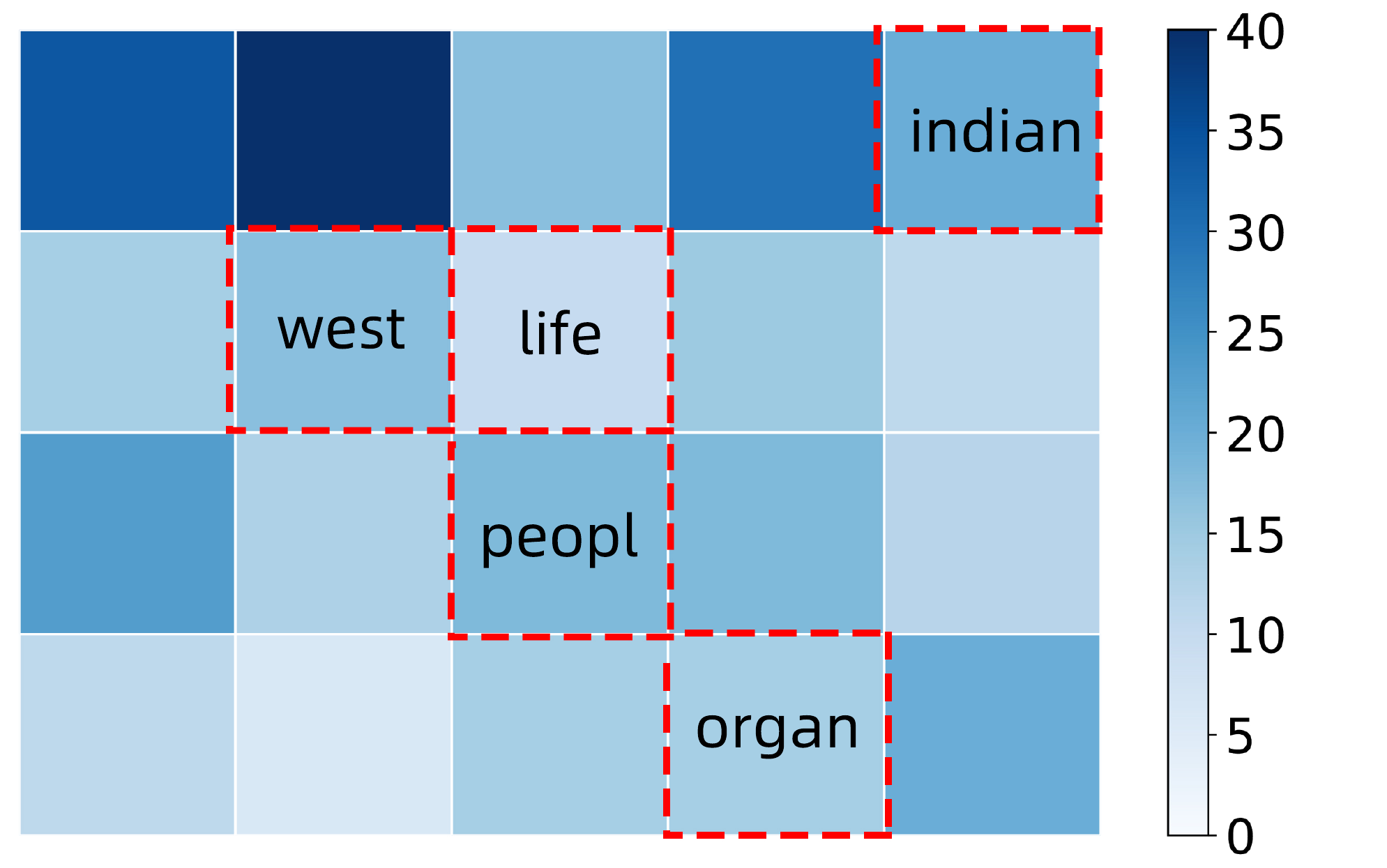}
			\centerline{\small{(i) term frequency}}
		\end{minipage}%
        \begin{minipage}[t]{0.49\linewidth}
			\includegraphics[width=\linewidth]{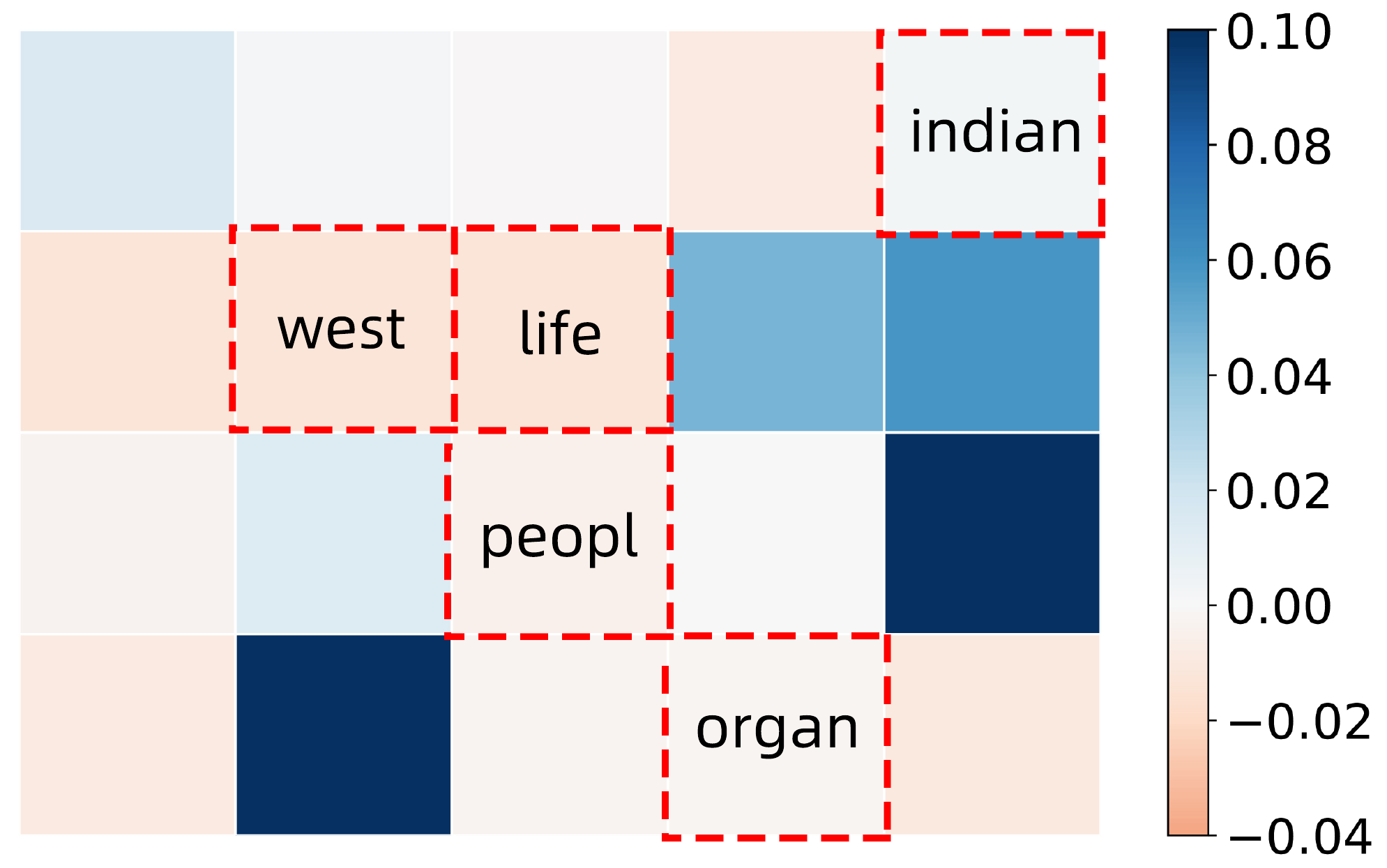}
	
			\centerline{\small{(ii) MAP promotion}}
		\end{minipage}%
    \end{minipage}%
	}%

	\subfigure[\normalsize{Top 20 terms assigned by QA4PRF (the red squares are different expansion terms compared with LL(ALL))}]{
	\begin{minipage}[t]{\linewidth}
		\begin{minipage}[t]{0.49\linewidth}
			\includegraphics[width=\linewidth]{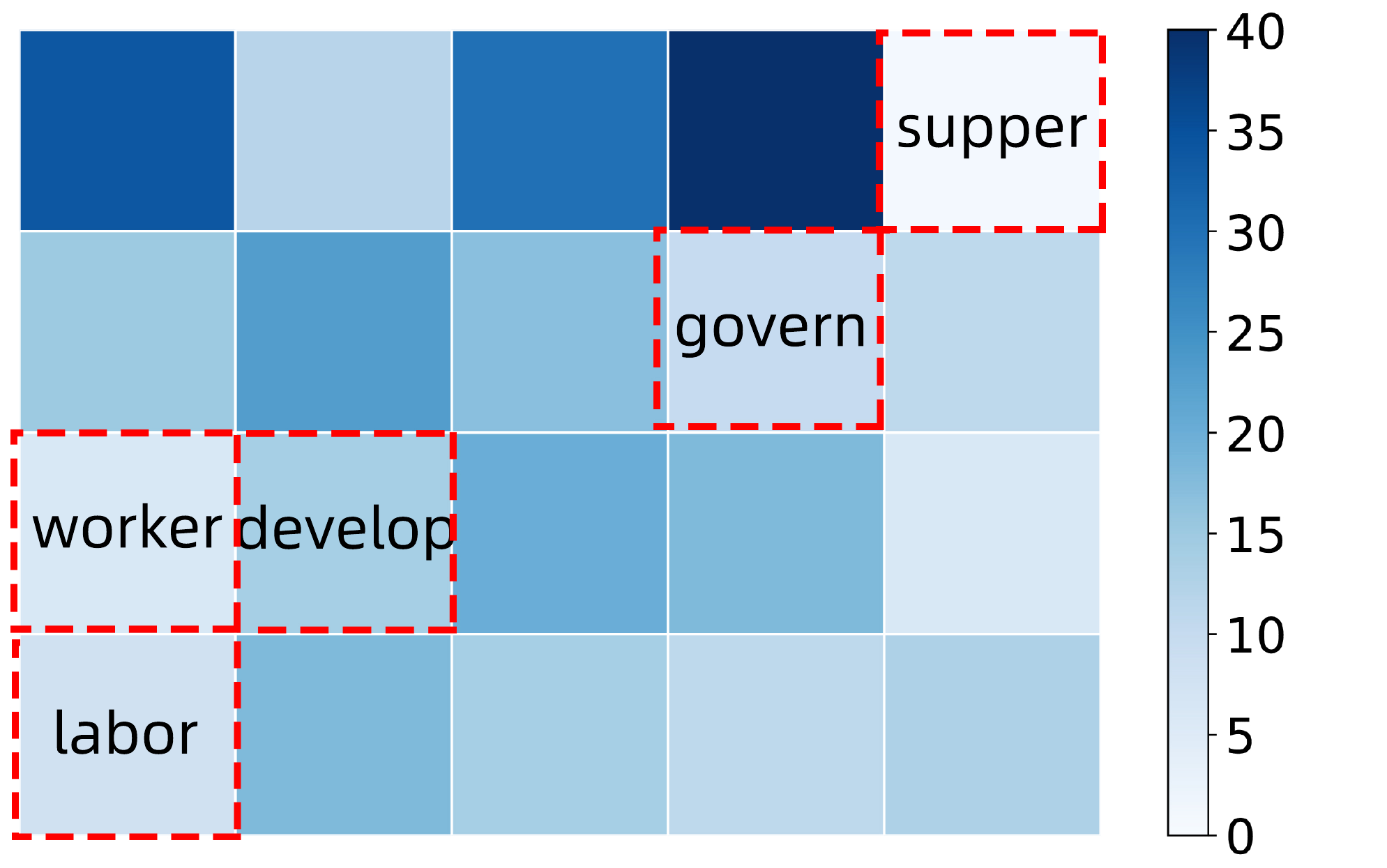}
			\centerline{\small{(i) term frequency}}
		\end{minipage}%
        \begin{minipage}[t]{0.49\linewidth}
			\includegraphics[width=\linewidth]{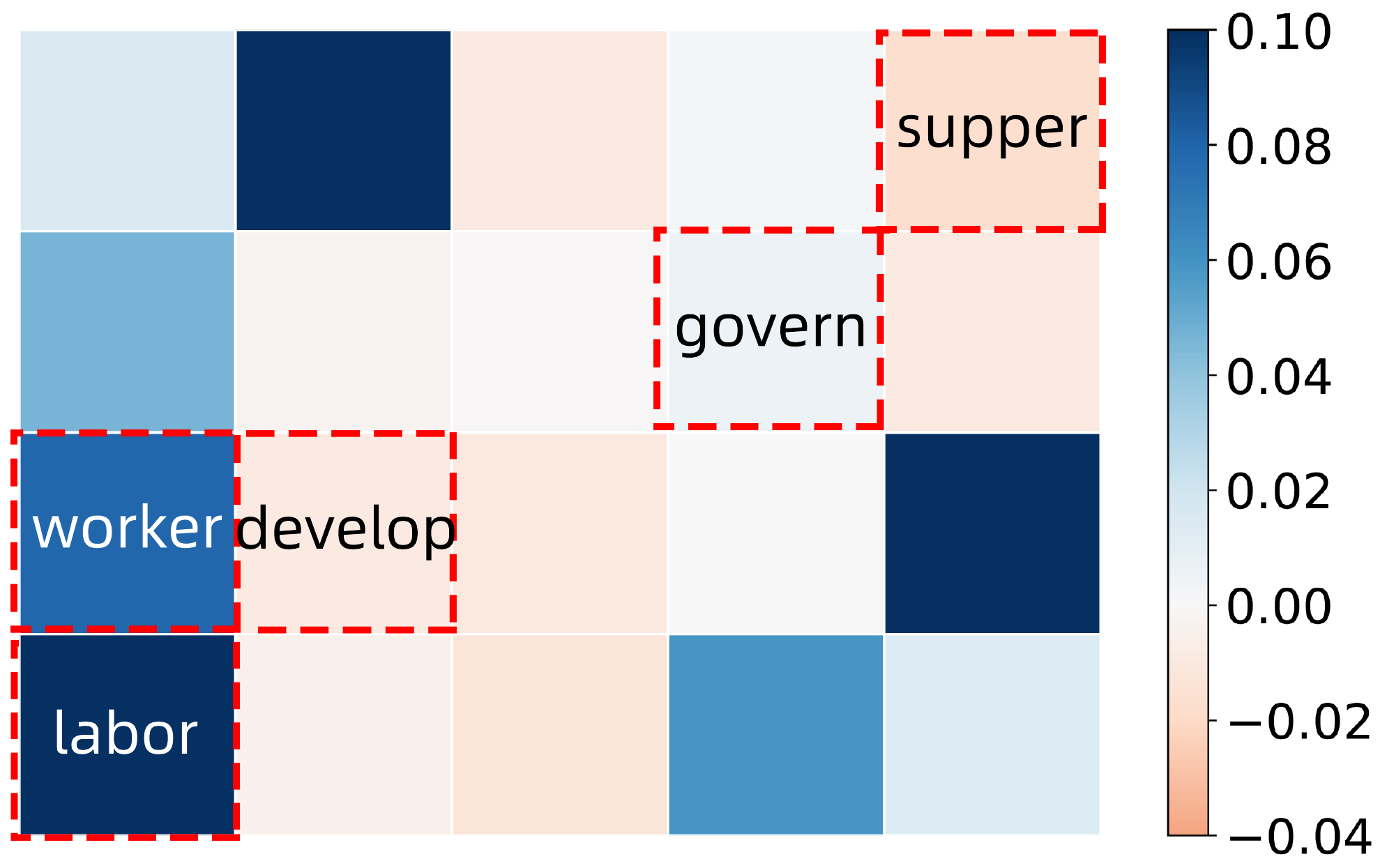}
			\centerline{\small{(ii) MAP promotion}}
		\end{minipage}%
    \end{minipage}%
	}%
	\caption{Top 20 terms selected by LL(ALL) and QA4PRF for query ``\textit{Modern Slavery}''. Each square represents a term and red squares highlight different terms between LL(ALL) and QA4PRF.}
	\label{pic:case}
\end{figure}

\subsection{Case Study (RQ4)}
To take a deep look into the characterization of selected terms by QA4PRF, we randomly pick two queries from TREC dataset and present the results of LL(ALL) and QA4PRF in Table~\ref{tab:case}. Terms in this section are all shown with their stemmers which has been discussed in Section~\ref{sec:exp-setup}.

Table~\ref{tab:case} shows top 20 expansion terms selected by LL(ALL) and QA4PRF respectively. Bold terms indicate the different terms selected by such two methods. As can be observed, both LL(ALL) and QA4PRF can provide several terms with high term frequency which are useful for retrieval. However, LL(ALL) may include noisy terms with high term frequency in top-retrieved documents as well. For example, ``\textit{life}'' and ``\textit{peopl}'' in query ``\textit{Modern Slavery}'', ``\textit{spokesman}'' and ``\textit{countri}'' in query ``\textit{Diplomatic Expulsion}'' are not informative. QA4PRF, on the contrary, is able to find useful and informative words for the given queries, e.g., ``\textit{worker}'', ``\textit{govern}'' and ``\textit{labor}'' represent main objects and reasons for ``\textit{Modern Slavery}'', ``\textit{russian}'', ``\textit{iraq}'' and ``\textit{attach}'' are the answers for ``\textit{where and why frequent Diplomatic Expulsion occur?}''.

To study such different selected terms by the two methods quantitatively, we take the query ``\textit{Modern Slavery}'' as an example and show the term frequency and MAP promotion of top 20 terms which selected by LL(ALL) and QA4PRF respectively in Figrue~\ref{pic:case}. Each square in Figure~\ref{pic:case} represents a term in Table~\ref{tab:case}. The expansion terms are arranged in descending order from left to right and top to bottom according to the score given by each model, corresponding to Table~\ref{tab:case}. Red squares highlight the different terms between two methods. Term frequency of a word is the total number of its occurrences in top 10 documents. It is obvious that LL(ALL) selects expansion terms with higher term frequency, compared to QA4PRF. According to the comparison of MAP promotion based on two models, we can see that, even though QA4PRF selects 5 different terms with less term frequency, such 5 terms lead to much better MAP improvement, compared to the ones selected by LL(ALL). This is achieved by managing to understand the content of top-retrieved documents to find expansion terms by QA4PRF.

\section{Conclusion}\label{sec:conclude}
In this work, we formulate pseudo relevance feedback (PRF) as a question answering (QA) task and propose a novel QA4-based framework for PRF called QA4PRF to utilize contextual information in documents, where the objective is to find some proper terms to expand the original query by utilizing contextual information. In QA4PRF framework, an attention-based pointer network is utilized to understand the top-retrieved documents in a human-interpretable way. Such a network is efficient and effective in capturing contextual interaction information among long word sequences in machine reading comprehension. Besides, we find that incorporating the traditional supervised learning methods, such as LambdaRank to make use of statistical PRF information further enhancing the performance of QA4PRF framework. Extensive experiments over three real-world datasets demonstrate that QA4PRF framework significantly outperforms all state-of-the-art PRF models. 

For future work, we plan to investigate reinforcement learning solutions to perform multi-step query reformulation in (pseudo) relevance feedback scenarios. In addition, extending QA4PRF for query reformulation in sponsored search would be the potential to improve the platform revenue when considering the auction competitiveness of each candidate term.






\bibliographystyle{IEEEtranN}
\bibliography{IEEEabrv, PRF}

\begin{thebibliography}{37}
\providecommand{\natexlab}[1]{#1}
\providecommand{\url}[1]{#1}
\csname url@samestyle\endcsname
\providecommand{\newblock}{\relax}
\providecommand{\bibinfo}[2]{#2}
\providecommand{\BIBentrySTDinterwordspacing}{\spaceskip=0pt\relax}
\providecommand{\BIBentryALTinterwordstretchfactor}{4}
\providecommand{\BIBentryALTinterwordspacing}{\spaceskip=\fontdimen2\font plus
\BIBentryALTinterwordstretchfactor\fontdimen3\font minus
  \fontdimen4\font\relax}
\providecommand{\BIBforeignlanguage}[2]{{%
\expandafter\ifx\csname l@#1\endcsname\relax
\typeout{** WARNING: IEEEtranN.bst: No hyphenation pattern has been}%
\typeout{** loaded for the language `#1'. Using the pattern for}%
\typeout{** the default language instead.}%
\else
\language=\csname l@#1\endcsname
\fi
#2}}
\providecommand{\BIBdecl}{\relax}
\BIBdecl

\bibitem[Carpineto and Romano(2012)]{carpineto2012survey}
C.~Carpineto and G.~Romano, ``A survey of automatic query expansion in
  information retrieval,'' \emph{Acm Computing Surveys (CSUR)}, vol.~44, 2012.

\bibitem[Rocchio(1971)]{rocchio1971relevance}
J.~J. Rocchio, ``Relevance feedback in information retrieval,'' \emph{The SMART
  retrieval system: experiments in automatic document processing}, 1971.

\bibitem[Salton and Buckley(1990)]{salton1990improving}
G.~Salton and C.~Buckley, ``Improving retrieval performance by relevance
  feedback,'' \emph{Journal of the American society for information science},
  1990.

\bibitem[Xu and Croft(1996)]{Xu:1996:QEU:243199.243202}
J.~Xu and W.~B. Croft, ``Query expansion using local and global document
  analysis,'' in \emph{Proc. of 19th SIGIR}.\hskip 1em plus 0.5em minus
  0.4em\relax ACM, 1996.

\bibitem[Lavrenko and Croft(2001)]{lavrenko2001relevance}
V.~Lavrenko and W.~B. Croft, ``Relevance-based language models,'' in
  \emph{Proc. of 24th SIGIR}.\hskip 1em plus 0.5em minus 0.4em\relax ACM, 2001.

\bibitem[Abdul-Jaleel et~al.(2004)Abdul-Jaleel, Allan, Croft, Diaz, Larkey, Li,
  Smucker, and Wade]{abdul2004umass}
N.~Abdul-Jaleel, J.~Allan, W.~B. Croft, F.~Diaz, L.~Larkey, X.~Li, M.~D.
  Smucker, and C.~Wade, ``Umass at trec 2004: Novelty and hard,'' 2004.

\bibitem[Roy et~al.(2019)Roy, Bhatia, and Mitra]{Roy2019Discriminative}
D.~Roy, S.~Bhatia, and M.~Mitra, ``Selecting discriminative terms for relevance
  model,'' in \emph{Proc. of 42nd SIGIR}.\hskip 1em plus 0.5em minus
  0.4em\relax ACM, 2019.

\bibitem[Zhai and Lafferty(2001)]{zhai2001model}
C.~Zhai and J.~Lafferty, ``Model-based feedback in the language modeling
  approach to information retrieval,'' in \emph{Proc. of 10th CIKM}.\hskip 1em
  plus 0.5em minus 0.4em\relax ACM, 2001.

\bibitem[Lv and Zhai(2014)]{lv2014revisiting}
Y.~Lv and C.~Zhai, ``Revisiting the divergence minimization feedback model,''
  in \emph{Proc. of 23rd CIKM}.\hskip 1em plus 0.5em minus 0.4em\relax ACM,
  2014.

\bibitem[Clinchant and Gaussier(2010)]{clinchant2010information}
S.~Clinchant and E.~Gaussier, ``Information-based models for ad hoc ir,'' in
  \emph{Proc. of 33rd SIGIR}.\hskip 1em plus 0.5em minus 0.4em\relax ACM, 2010.

\bibitem[Montazeralghaem et~al.(2017)Montazeralghaem, Zamani, and
  Shakery]{montazeralghaem2017term}
A.~Montazeralghaem, H.~Zamani, and A.~Shakery, ``Term proximity constraints for
  pseudo-relevance feedback,'' in \emph{Proc. of 40th SIGIR}.\hskip 1em plus
  0.5em minus 0.4em\relax ACM, 2017.

\bibitem[Montazeralghaem et~al.(2018)Montazeralghaem, Zamani, and
  Shakery]{montazeralghaem2018theoretical}
A.~Montazeralghaem, H.~Zamani, and Shakery, ``Theoretical analysis of
  interdependent constraints in pseudo-relevance feedback,'' in \emph{Proc. of
  41st SIGIR}.\hskip 1em plus 0.5em minus 0.4em\relax ACM, 2018.

\bibitem[Zamani et~al.(2016)Zamani, Dadashkarimi, Shakery, and
  Croft]{zamani2016pseudo}
H.~Zamani, J.~Dadashkarimi, A.~Shakery, and W.~B. Croft, ``Pseudo-relevance
  feedback based on matrix factorization,'' in \emph{Proc. of 25th CIKM}, 2016.

\bibitem[Cao et~al.(2008)Cao, Nie, Gao, and Robertson]{cao2008selecting}
G.~Cao, J.-Y. Nie, J.~Gao, and S.~Robertson, ``Selecting good expansion terms
  for pseudo-relevance feedback,'' in \emph{Proc. of 31st SIGIR}.\hskip 1em
  plus 0.5em minus 0.4em\relax ACM, 2008.

\bibitem[Roy et~al.(2016)Roy, Paul, Mitra, and Garain]{roy2016using}
D.~Roy, D.~Paul, M.~Mitra, and U.~Garain, ``Using word embeddings for automatic
  query expansion,'' \emph{arXiv:1606.07608}, 2016.

\bibitem[Kuzi et~al.(2016)Kuzi, Shtok, and Kurland]{kuzi2016query}
S.~Kuzi, A.~Shtok, and O.~Kurland, ``Query expansion using word embeddings,''
  in \emph{Proc. of 25th CIKM}.\hskip 1em plus 0.5em minus 0.4em\relax ACM,
  2016.

\bibitem[Rajpurkar et~al.(2016)Rajpurkar, Zhang, Lopyrev, and
  Liang]{rajpurkar2016squad}
P.~Rajpurkar, J.~Zhang, K.~Lopyrev, and P.~Liang, ``Squad: 100,000+ questions
  for machine comprehension of text,'' \emph{arXiv:1606.05250}, 2016.

\bibitem[Vaswani et~al.(2017)Vaswani, Shazeer, Parmar, Uszkoreit, Jones, Gomez,
  Kaiser, and Polosukhin]{vaswani2017attention}
A.~Vaswani, N.~Shazeer, N.~Parmar, J.~Uszkoreit, L.~Jones, A.~N. Gomez,
  {\L}.~Kaiser, and I.~Polosukhin, ``Attention is all you need,'' in
  \emph{Proc. of 30th NIPS}, 2017.

\bibitem[Seo et~al.(2017)Seo, Kembhavi, Farhadi, and
  Hajishirzi]{seo2016bidirectional}
M.~Seo, A.~Kembhavi, A.~Farhadi, and H.~Hajishirzi, ``Bidirectional attention
  flow for machine comprehension,'' 2017.

\bibitem[Yu et~al.(2018)Yu, Dohan, Luong, Zhao, Chen, Norouzi, and
  Le]{yu2018qanet}
A.~W. Yu, D.~Dohan, M.-T. Luong, R.~Zhao, K.~Chen, M.~Norouzi, and Q.~V. Le,
  ``Qanet: Combining local convolution with global self-attention for reading
  comprehension,'' 2018.

\bibitem[Vinyals et~al.(2015)Vinyals, Fortunato, and
  Jaitly]{vinyals2015pointer}
O.~Vinyals, M.~Fortunato, and N.~Jaitly, ``Pointer networks,'' in \emph{Proc.
  of 28th NIPS}, 2015.

\bibitem[Qiu et~al.(2018)Qiu, Zhou, Qu, Zhang, Li, Rong, Ru, Qian, Tu, and
  Yu]{qiu2018qa4ie}
L.~Qiu, H.~Zhou, Y.~Qu, W.~Zhang, S.~Li, S.~Rong, D.~Ru, L.~Qian, K.~Tu, and
  Y.~Yu, ``Qa4ie: A question answering based framework for information
  extraction,'' in \emph{Proc. of 17th ISWC}, 2018.

\bibitem[Group(2017)]{group2017rnet}
\BIBentryALTinterwordspacing
N.~L.~C. Group, ``R-net: Machine reading comprehension with self-matching
  networks,'' May 2017. [Online]. Available:
  \url{https://www.microsoft.com/en-us/research/publication/mcr/}
\BIBentrySTDinterwordspacing

\bibitem[Raza et~al.(2018)Raza, Mokhtar, and Noraziah]{raza2018survey}
M.~A. Raza, R.~Mokhtar, and A.~Noraziah, ``A survey of statistical approaches
  for query expansion,'' \emph{Knowledge and Information Systems}, 2018.

\bibitem[Burges(2010)]{burges2010ranknet}
C.~J. Burges, ``From ranknet to lambdarank to lambdamart: An overview,''
  \emph{Learning}, vol.~11, no. 23-581, 2010.

\bibitem[Lv and Zhai(2009)]{lv2009comparative}
Y.~Lv and C.~Zhai, ``A comparative study of methods for estimating query
  language models with pseudo feedback,'' in \emph{Proc. of 18th CIKM}, 2009.

\bibitem[Lucchese et~al.(2018)Lucchese, Nardini, Perego, Trani, and
  Venturini]{lucchese2018efficient}
C.~Lucchese, F.~M. Nardini, R.~Perego, R.~Trani, and R.~Venturini, ``Efficient
  and effective query expansion for web search,'' in \emph{Proc. of 27th
  CIKM}.\hskip 1em plus 0.5em minus 0.4em\relax ACM, 2018.

\bibitem[Mikolov et~al.(2018)Mikolov, Grave, Bojanowski, Puhrsch, and
  Joulin]{mikolov2018advances}
T.~Mikolov, E.~Grave, P.~Bojanowski, C.~Puhrsch, and A.~Joulin, ``Advances in
  pre-training distributed word representations,'' in \emph{Proc. of 11th
  LREC}, 2018.

\bibitem[Chen et~al.(2017)Chen, Fisch, Weston, and Bordes]{chen2017reading}
D.~Chen, A.~Fisch, J.~Weston, and A.~Bordes, ``Reading wikipedia to answer
  open-domain questions,'' 2017.

\bibitem[Wang et~al.(2017)Wang, Yang, Wei, Chang, and Zhou]{wang2017gated}
W.~Wang, N.~Yang, F.~Wei, B.~Chang, and M.~Zhou, ``Gated self-matching networks
  for reading comprehension and question answering,'' in \emph{Proc. of 55th
  ACL}, 2017.

\bibitem[Loper and Bird(2002)]{loper2002nltk}
E.~Loper and S.~Bird, ``Nltk: the natural language toolkit,'' \emph{arXiv
  preprint cs/0205028}, 2002.

\bibitem[Robertson et~al.(1995)Robertson, Walker, Jones, Hancock-Beaulieu,
  Gatford, et~al.]{robertson1995okapi}
S.~E. Robertson, S.~Walker, S.~Jones, M.~M. Hancock-Beaulieu, M.~Gatford
  \emph{et~al.}, ``Okapi at trec-3,'' \emph{Nist Special Publication Sp}, vol.
  109, 1995.

\bibitem[Robertson and Walker(1994)]{robertson1994some}
S.~E. Robertson and S.~Walker, ``Some simple effective approximations to the
  2-poisson model for probabilistic weighted retrieval,'' in \emph{Proc. of
  17th SIGIR}.\hskip 1em plus 0.5em minus 0.4em\relax ACM, 1994.

\bibitem[Diaz et~al.(2016)Diaz, Mitra, and Craswell]{diaz2016query}
F.~Diaz, B.~Mitra, and N.~Craswell, ``Query expansion with locally-trained word
  embeddings,'' 2016.

\bibitem[Collins-Thompson(2009)]{collins2009reducing}
K.~Collins-Thompson, ``Reducing the risk of query expansion via robust
  constrained optimization,'' in \emph{Proc. of 18th CIKM}.\hskip 1em plus
  0.5em minus 0.4em\relax ACM, 2009.

\bibitem[Wilcoxon(1992)]{wilcoxon1992individual}
F.~Wilcoxon, ``Individual comparisons by ranking methods,'' in
  \emph{Breakthroughs in statistics}.\hskip 1em plus 0.5em minus 0.4em\relax
  Springer, 1992.

\bibitem[Clinchant and Gaussier(2013)]{clinchant2013theoretical}
S.~Clinchant and E.~Gaussier, ``A theoretical analysis of pseudo-relevance
  feedback models,'' in \emph{Proc. of 4th ICTIR}.\hskip 1em plus 0.5em minus
  0.4em\relax ACM, 2013.

\end{thebibliography}

\begin{IEEEbiography}[{\includegraphics[width=1in,height=1.25in,clip,keepaspectratio]{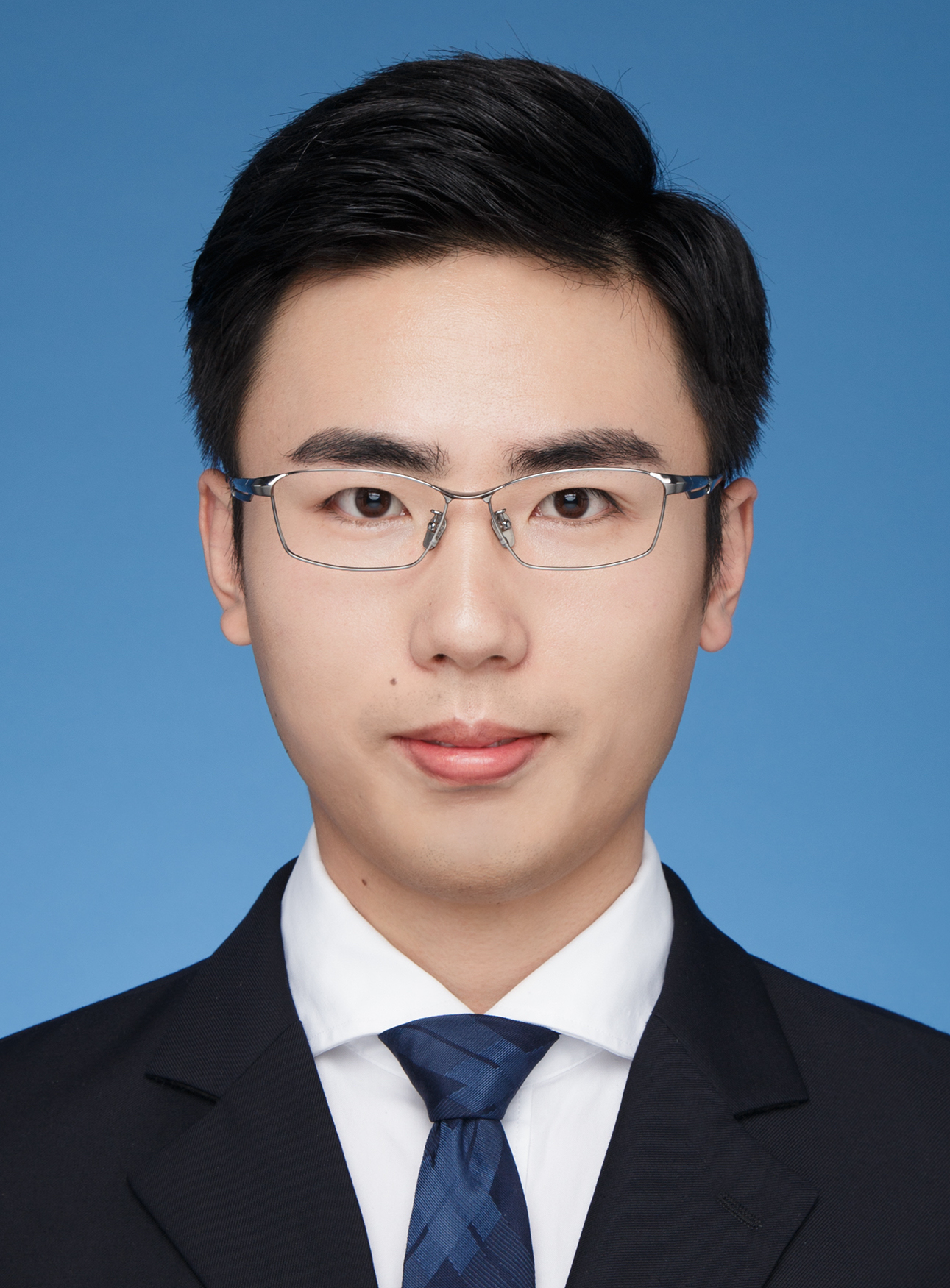}}]{Handong Ma} is now a PhD candidate at department of computer science in Shanghai Jiaotong University. He earned his master degree from Columbia University Medical Center department of Biomedical Informatics in 2015. Afterwards he worked as data analyst in Pfizer New York and data scientist in Columbia University Medical Center.
Handong ma’s research interests includes Natural Language Processing in medical domain, especially the secondary use of Electronic Medical Record (EMR). He is also working in building machine learning models for real-world medical use including data extraction, clinical decision support etc. He has published multiple research papers in biomedical informatics journals such as JBI and JMIR together with other medical journals covering multiple domains. 
\end{IEEEbiography}

\begin{IEEEbiography}[{\includegraphics[width=1in,height=1.25in,clip,keepaspectratio]{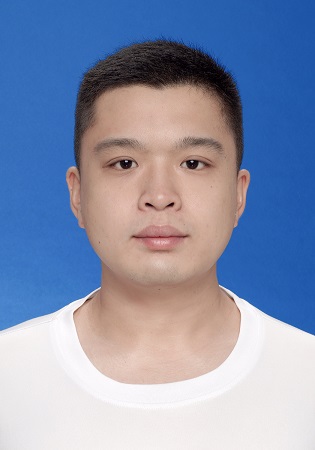}}]{Jiawei Hou} is a master student in Apex Data \& Knowledge Management Lab, Shanghai Jiao Tong University, advised by Prof. Weinan Zhang and Prof.Yong Yu. He received his B.E. degree from the Department of Computer Science,  Shanghai Jiao Tong University in 2018. His research interests include data mining and information retrieval.
\end{IEEEbiography}

\begin{IEEEbiography}[{\includegraphics[width=1in,height=1.25in,clip,keepaspectratio]{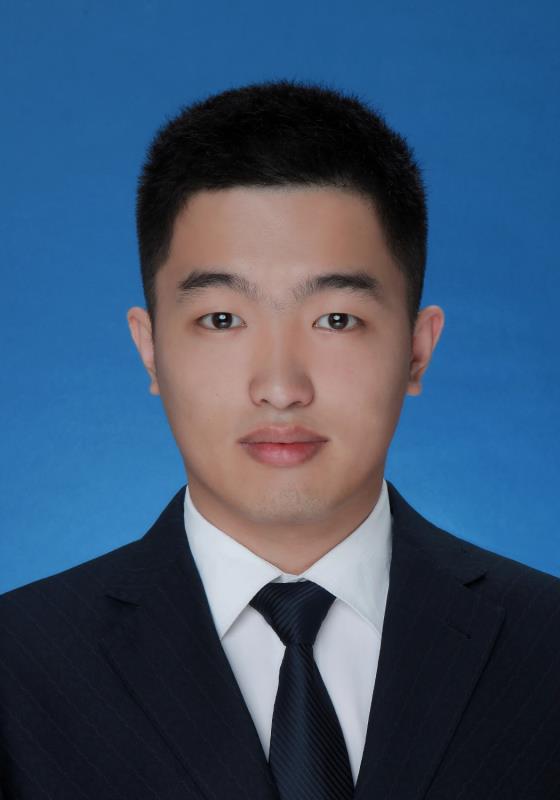}}]{Chenxu Zhu} is a computer science master student in Apex Data \& Knowledge Management Lab, Department of Computer Science, Shanghai Jiao Tong University, advised by Prof.Yong Yu and Prof. Weinan Zhang. He received his B.E. degree from Shanghai Jiao Tong University in 2020. His research interests include data mining, machine learning, reinforcement learning and recommender system.
\end{IEEEbiography}

\begin{IEEEbiography}[{\includegraphics[width=1in,height=1.25in,clip,keepaspectratio]{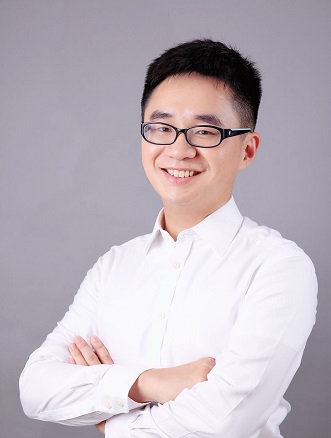}}]{Weinan Zhang} is now a tenure-track associate professor at Shanghai Jiao Tong University. His research interests include  reinforcement learning, deep learning and data science with various real-world applications of recommender systems, search engines, text mining \& generation, knowledge graphs, game AI etc. He has published over 80 research papers on international conferences and journals and has been serving as a (senior) PC member at ICML, NeurIPS, ICLR, KDD, AAAI, IJCAI, SIGIR etc. and a reviewer at JMLR, TOIS, TKDE, TIST etc. 
\end{IEEEbiography}

\begin{IEEEbiography}[{\includegraphics[width=1in,height=1.25in,clip,keepaspectratio]{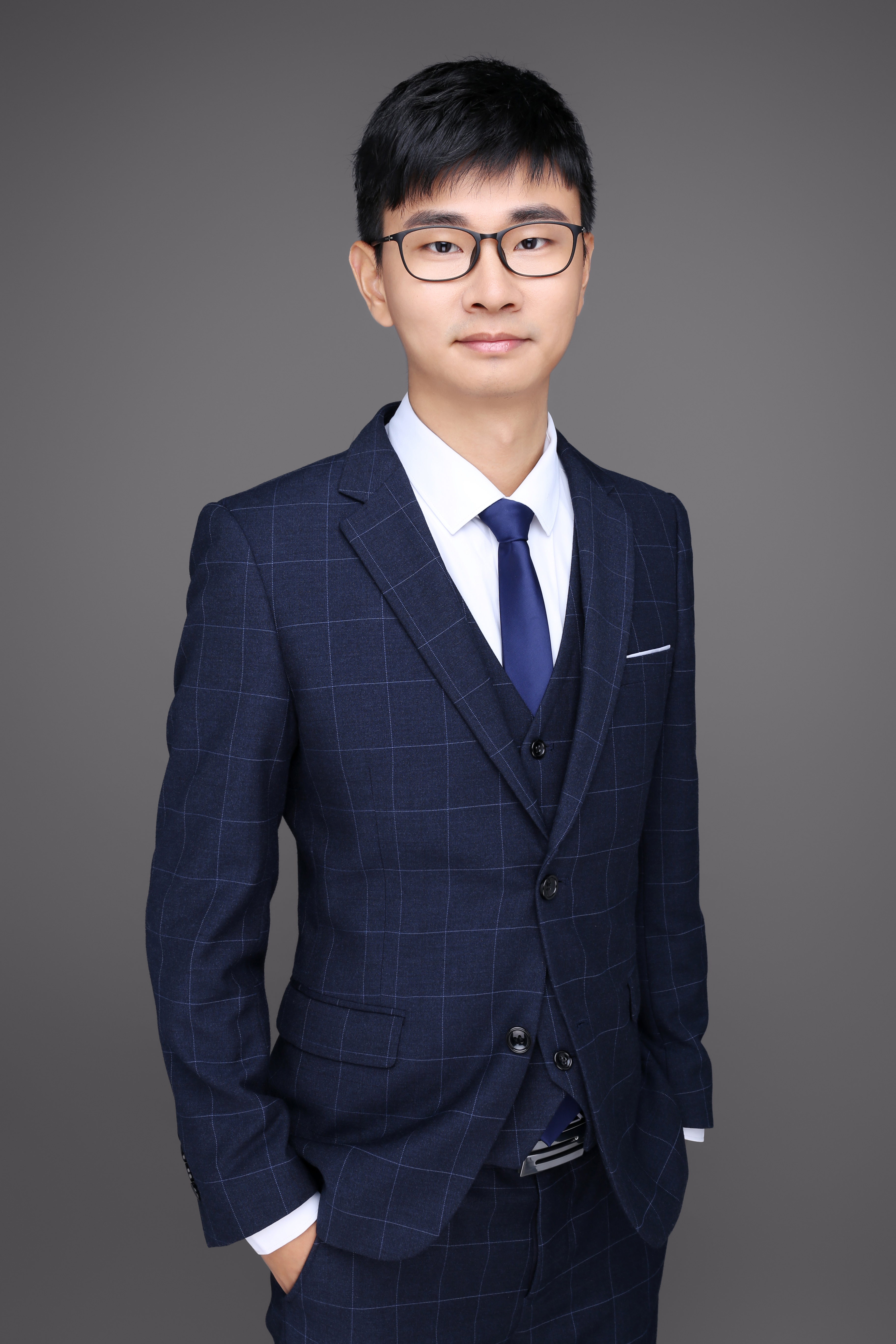}}]{Ruiming Tang} is a senior researcher in recommendation and search project team, Huawei Noah’s Ark Lab. He joint Noah’s Ark Lab in 2014. His research topics include recommender system, deep learning, reinforcement learning, AutoML, Graph Neural Network and etc. He published multiple research works on top-tier conferences and journals, on the topic of recommender system, such as WWW, IJCAI, SIGIR, RecSys, AAAI, TOIS, WSDM, KDD, CIKM. 
Before joining Huawei, Ruiming received his Ph.D. degree in Computer Science from National University of Singapore (NUS) in 2014 and received his Bachelor degree in Computer Science from Northeastern University in China (NEU) in 2009.
\end{IEEEbiography}

\begin{IEEEbiography}[{\includegraphics[width=1in,height=1.25in,clip,keepaspectratio]{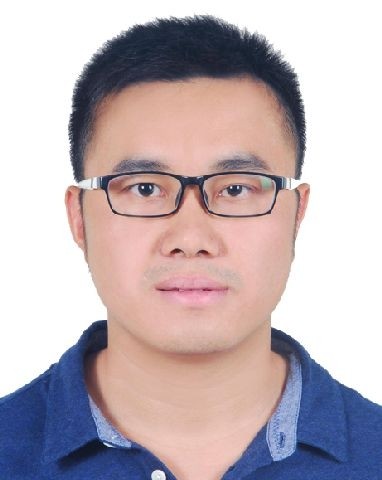}}]{Jincai Lai} is a researcher in recommendation and search project team, Huawei Noah’s Ark Lab.  He joint Noah’s Ark Lab in 2018. His research topics include recommender system,  deep  learning, information retrival and etc. He received the B.S in Communication Engineering and M.S in Computer Science from Beijing University of Posts and Telecommunications(BUPT) in 2015 and 2018. 
\end{IEEEbiography}

\begin{IEEEbiography}[{\includegraphics[width=1in,height=1.25in,clip,keepaspectratio]{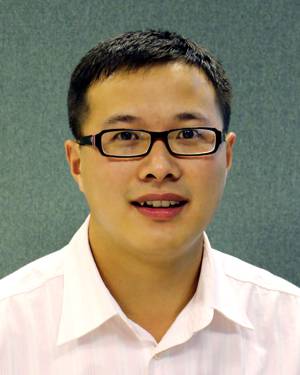}}]{Jieming Zhu} received the PhD degree from the Department of
Computer Science and Engineering, The Chinese University of Hong Kong, in 2016.
He is currently a researcher at Huawei Noah’s Ark Lab. His research interests include recommender systems, multimodal learning, natural language processing, and log analysis.
\end{IEEEbiography}

\begin{IEEEbiography}[{\includegraphics[width=1in,height=1.25in,clip,keepaspectratio]{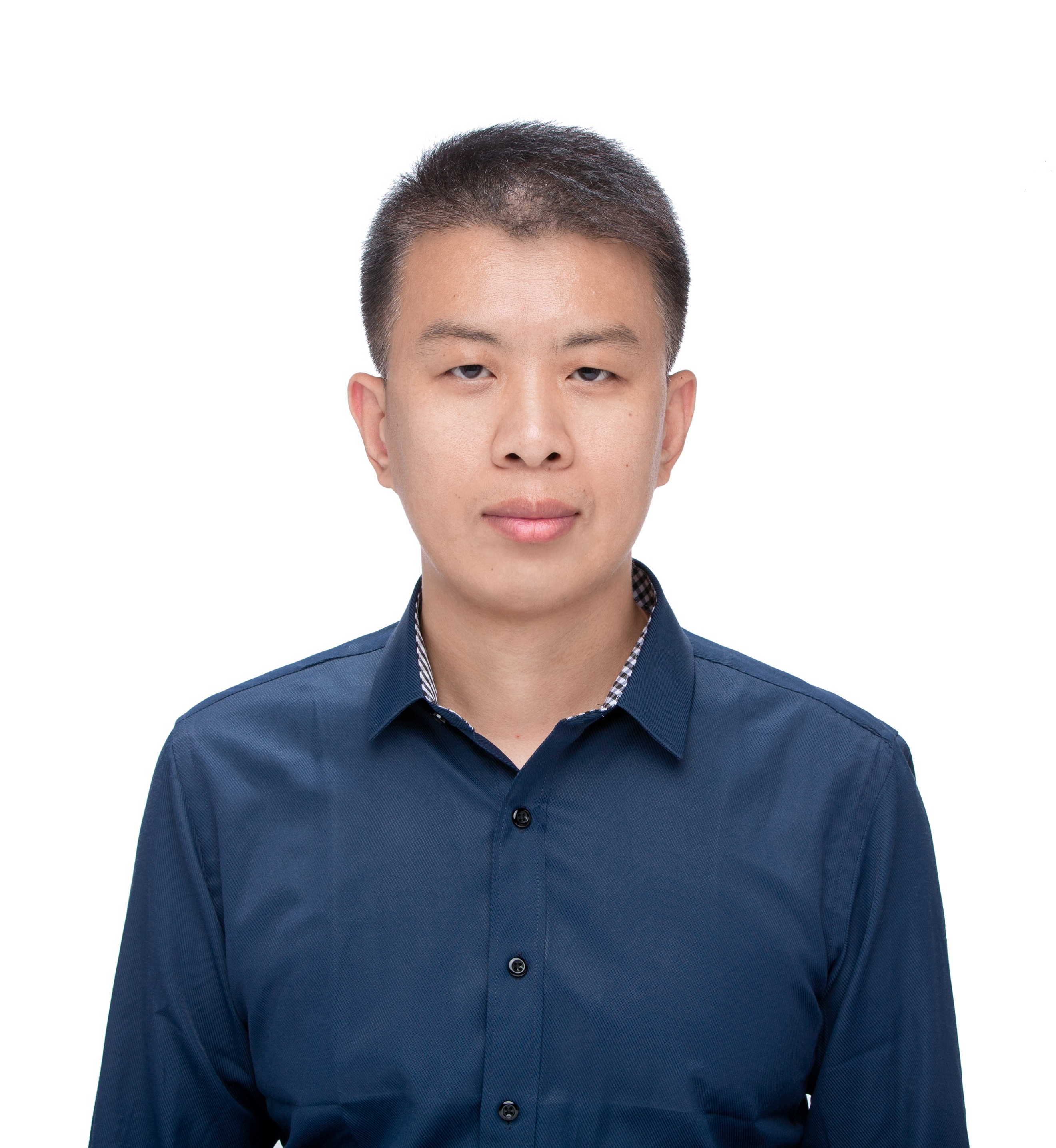}}]{Xiuqiang He} is currently the director of Recommendation \& Search Lab and Principal Researcher in Huawei Noah’s Ark Lab. He received the B.S. and M.S. degrees from the Department of Computer Science at Xi’an Jiaotong University, in 2003 and 2006, respectively, and the Ph.D. degree from the Department of Computer Science at the Hong Kong University of Science and Technology, in 2010. His research interests include machine learning algorithms in the area of recommendation and search.
\end{IEEEbiography}

\begin{IEEEbiography}[{\includegraphics[width=1in,height=1.25in,clip,keepaspectratio]{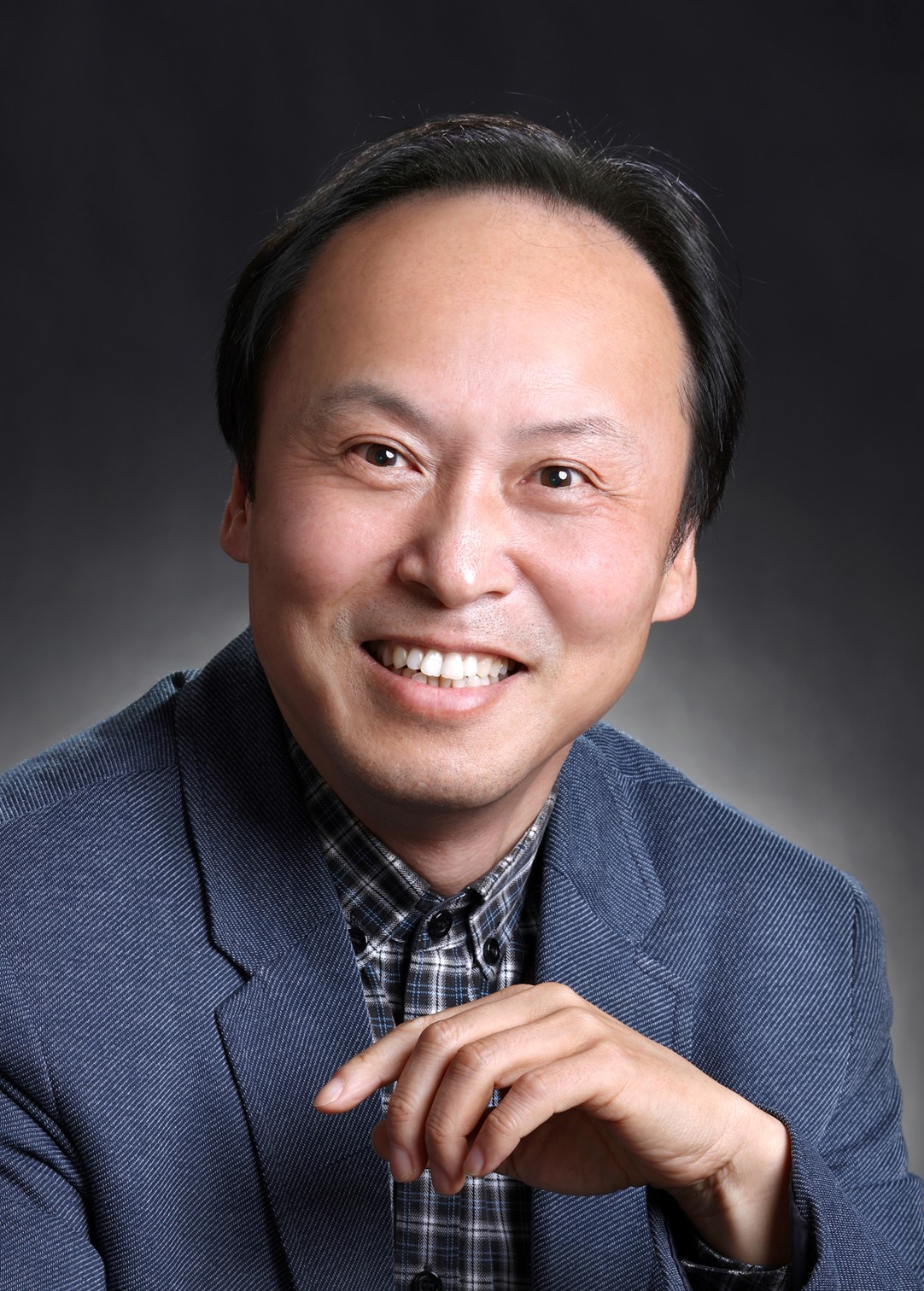}}]{Yong Yu} is a professor in Department of Computer Science in Shanghai Jiao Tong University. His research interests include information
systems, web search, data mining and machine
learning. He has published over 200 papers and
served as PC member of several conferences
including WWW, RecSys and a dozen of other
related conferences (e.g., NIPS, ICML, SIGIR,
ISWC etc.) in these fields.
\end{IEEEbiography}

\EOD

\end{document}